\documentclass[reprint, aps, amsmath, amssymb, pra, 10pt, floatfix, notitlepage]{revtex4-2}

\usepackage[english]{babel}   
\usepackage[pdftex]{graphicx}
\usepackage{amsmath, color, colordvi}
\usepackage{dcolumn}
\usepackage{bm}
\definecolor{navyblue}{rgb}{0.0, 0.0, 0.5}

\usepackage{hyperref}
\usepackage[draft]{changes}
\hypersetup{
    unicode=false,          
    pdftoolbar=false,        
    pdfmenubar=true,        
    pdffitwindow=false,     
    pdfstartview={FitH},    
    pdftitle={},    
    pdfauthor={},     
    pdfsubject={},   
    pdfcreator={},   
    pdfproducer={}, 
    pdfkeywords={many-body localization} {non-ergodic dynamics}{bosons}{level statistics}{thermalization}, 
    pdfnewwindow=true,      
    colorlinks=true,       
    linkcolor=blue,          
    citecolor=blue,        
    filecolor=magenta,      
    urlcolor=blue           
}

\usepackage[protrusion=true, expansion=true]{microtype} 
\usepackage{braket}
\usepackage[per-mode=symbol, separate-uncertainty, range-phrase=--, range-units=single]{siunitx}
\usepackage{multirow}

\usepackage{float}

\DeclareMathOperator{\Tr}{Tr}

\usepackage{mathtools}

\newcommand{\aop}{\hat a}
\newcommand{\nop}{\hat n}
\newcommand{\ntot}{\hat N}
\newcommand{\hop}{\hat H}

\newcommand{\dens}{\hat\rho}

\newcommand{\oop}{\hat O}
\newcommand{\lind}{\mathcal{L}}
\newcommand{\rhoss}{\dens_{\rm ss}}
\newcommand{\bop}{\hat b}
\newcommand{\cop}{\hat c}
\newcommand{\dop}{\hat d}
\newcommand{\lev}{d}

\newcommand{\spop}{\hat\sigma_+}
\newcommand{\smop}{\hat\sigma_-}
\newcommand{\sxop}{\hat\sigma_x}
\newcommand{\szop}{\hat\sigma_z}

\newcommand{\xiop}{\hat\xi}

\let\vec\oldvec
\newcommand{\vec}{\mathbf}
\newcommand{\mat}{\textsf}

\DeclareMathOperator{\im}{Im}
\DeclareMathOperator{\pv}{p{.}v{.}}

\makeatletter
\def\bbl@set@language#1{%
  \edef\languagename{%
    \ifnum\escapechar=\expandafter`\string#1\@empty
    \else\string#1\@empty\fi}%
  \@ifundefined{babel@language@alias@\languagename}{}{%
    \edef\languagename{\@nameuse{babel@language@alias@\languagename}}%
  }%
  \select@language{\languagename}%
  \expandafter\ifx\csname date\languagename\endcsname\relax\else
    \if@filesw
      \protected@write\@auxout{}{\string\select@language{\languagename}}%
      \bbl@for\bbl@tempa\BabelContentsFiles{%
        \addtocontents{\bbl@tempa}{\xstring\select@language{\languagename}}}%
      \bbl@usehooks{write}{}%
    \fi
  \fi}
\newcommand{\DeclareLanguageAlias}[2]{%
  \global\@namedef{babel@language@alias@#1}{#2}%
}
\makeatother
\DeclareLanguageAlias{en}{english}

\makeatletter
\def\MT@warn@unknown{}
\makeatother

\hfuzz=\maxdimen
\newcount\hbadness
\newdimen\hfuzz

\begin{document}

\title{Collective bosonic effects in an array of transmon devices}

\author{Tuure~Orell$^1$}
\author{Maximilian~Zanner$^{2,3}$}
\author{Mathieu~L.~Juan$^4$}
\author{Aleksei~Sharafiev$^{2,3}$}
\author{Romain~Albert$^{2,3}$}
\author{Stefan~Oleschko$^{2,3}$}
\author{Gerhard~Kirchmair$^{2,3}$}
\author{Matti~Silveri$^1$}

\affiliation{$^1$Nano and Molecular Systems Research Unit, University of Oulu, 90014 Oulu, Finland, \\
$^2$Institute for Experimental Physics, University of Innsbruck, A-6020 Innsbruck, Austria\\
$^3$Institute for Quantum Optics and Quantum Information, Austrian Academy of Sciences, A-6020 Innsbruck, Austria\\
$^4$Institut Quantique and Département de Physique, Université de Sherbrooke, Sherbrooke J1K2R1 Québec, Canada
}

\date{\today}
\begin{abstract}
    Multiple emitters coherently interacting with an electromagnetic mode give rise to collective effects such as correlated decay and coherent exchange interaction, depending on the separation of the emitters. By diagonalizing the effective non-Hermitian many-body Hamiltonian we reveal the complex-valued eigenvalue spectrum encoding the decay and interaction characteristics. We show that there are significant differences in the emerging effects for an array of interacting anharmonic oscillators compared to those of two-level systems and harmonic oscillators. The bosonic decay rate of the most superradiant state increases linearly as a function of the filling factor and exceeds that of two-level systems in magnitude. Furthermore, with bosonic systems, dark states are formed at each filling factor. These are in strong contrast with two-level systems, where the maximal superradiance is observed at half filling and with larger filling factors superradiance diminishes and no dark states are formed. As an experimentally relevant setup of bosonic waveguide QED, we focus on arrays of transmon devices embedded inside a rectangular waveguide.
    Specifically, we study the setup of two transmon pairs realized experimentally in M. Zanner et al., arXiv.2106.05623 (2021), and show that it is necessary to consider transmons as bosonic multilevel emitters to accurately recover correct collective effects for the higher excitation manifolds.   
\end{abstract}
\maketitle

\section{Introduction}
Electromagnetic field inside a waveguide acts as a collective environment for quantum emitters embedded therein~\cite{Gu2017, Sheremet2021}. Coupling to a continuum of radiation modes leads to the emergence of collective states. Frequencies or positioning of individual emitters controls the relative phases, which determines whether the collective states are rapidly decaying superradiant states, or slowly decaying subradiant states~\cite{lalumiere2013, vanLoo2013}.

Superconducting circuits offer multiple advantages over atoms~\cite{Goban2015, Vahala2003, Chang2013} as emitters. One has larger control over the system parameters, which can be adjusted to desired values during fabrication, and even controlled in-situ during experiments. The frequencies of superconducting circuits are flux-tunable~\cite{Koch2007, Fitzpatrick2017, Gargiulo_21}, which can be utilized very efficiently in rectangular waveguides, which have a cutoff frequency determined by their dimensions. Radiation below the cutoff cannot propagate, and emitters below the cutoff are effectively secluded from the system. Thus, superconducting-circuit based emitters can be easily decoupled from the waveguide by tuning their frequencies below the cutoff.

This controllability combined with the unitary on-site control~\cite{Zanner2021} implies that waveguide QED based on superconducting circuits has a high potential in many applications, ranging from the simulation of dynamics of interacting quantum systems~\cite{Gonzalez-Tudela2015} to open quantum information processing and computation~\cite{Paulisch2016}, and even modeling light-harvesting~\cite{Celardo2014} and non-Markovian effects~\cite{Laakso2014}. Further, the space inside a rectangular waveguide makes it possible to realize three dimensional emitter constructions using the superconducting circuits~\cite{Hacohen15, Dalmonte2015, Kou18}.

\begin{figure}[t!]
    \centering
    \includegraphics[width=1.0\linewidth]
    {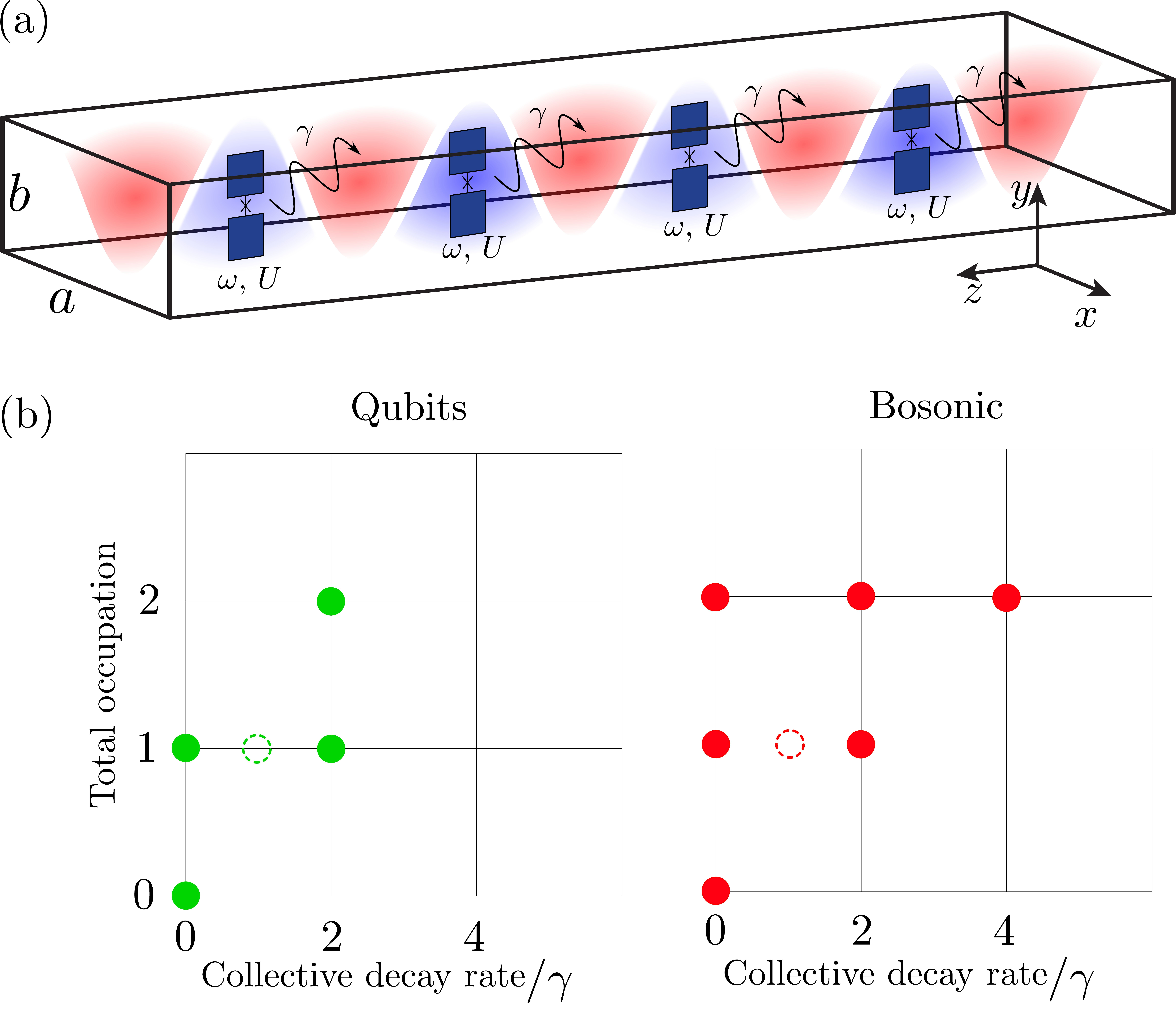}
    \caption{(a)~Transmons with frequency~$\omega$ and anharmonicity~$U$ coupled to the waveguide field with rate~$\gamma$. Waveguide is rectangular with width~$a$ and height~$b$. (b)~Total occupation and decay rates of the collective states in a two-site system, where the sites are either qubits (green) or harmonic oscillators (red). Bosonic statistics enhance the decay rates of the bright states, and enable dark states beyond half filling. Dashed circles depict states without collective effects.}
    \label{fig:circuit}
\end{figure}

Research on the collective phenomena has widely focused on two-level systems both theoretically and experimentally~\cite{lalumiere2013, Mirhosseini2019, Molmer2019, Molmer2020, Albrecht2019, Masson2020}. Collectively decaying two-level systems are known to exhibit the famous Dicke superradiance~\cite{Dicke1954} and superradiant radiation burst~\cite{Slama2007, Scheibner2007, masson2021universality}. Superconducting circuits are often referred to as qubits, but in reality they are more accurately described as quantum multilevel systems. In this work we consider an array of transmons in a rectangular waveguide, as sketched in Fig.~\ref{fig:circuit}(a). A transmon is an anharmonic oscillator~\cite{Koch2007}, and the anharmonicity acts as an on-site interaction between the excitations. It reduces the energies of multiply excited states, so that they are detuned from the transition between the ground state and the first excited state. Thus, a transmon can be seen as an intermediate between a qubit, which it resembles for large anharmonicity, and a harmonic oscillator in the opposite limit. The excitations in a transmon can be interpreted as bosons, and an array of transmons can be described accurately with the Bose--Hubbard model with attractive interactions~\cite{Hacohen15, Roushan17, Ma2019, Orell2019, Mansikkamaki2021}.
 
Bosonic statistics has a large impact on the collective decay of the system, and the difference becomes visible already with two sites one wavelength apart from each other, as depicted in Fig.~\ref{fig:circuit}(b).~This arises from the larger many-body Hilbert space of bosonic systems, as opposed to qubits. In qubit systems, the decay rates of the collective states start to decrease after half-filling, and dark states do not exist beyond that. In bosonic systems, on the other hand, the decay rates increase linearly, and the dark states can exist with any occupation.
  
This paper is organized in a following way. In Sec.~\ref{sec:waveguideQED} we describe the transmon system and its interaction with the electromagnetic field inside a rectangular waveguide. The dynamics of the transmons are described by a collective master equation, and the two cases where the transmon frequencies are tuned above and below the cutoff frequency of the waveguide are discussed. In Sec.~\ref{sec:collective} we study the collective effects of an array of bosonic sites and compare them to the better known array of two-level systems by studying the eigenvalues of non-Hermitian effective Hamiltonians. In particular, we find that the bosonic statistics enhance the superradiance. In Sec.~\ref{sec:interplay} we discuss the effect of direct coupling between the emitters by considering a system where pairs of capacitively coupled transmons are evenly distributed along the waveguide. This leads to an intriguing internal structure: local dark and bright states for each pair. The bright states further combine to global dark and bright states, extending throughout the entire system. Finally, we discuss more thoroughly the special case of two pairs, which was studied experimentally in Ref.~\cite{Zanner2021}. In Sec.~\ref{sec:spectroscopy} we discuss possible ways to probe the collective effects and states in transmon systems. We study the superradiant burst in bosonic systems, and use the two-pair setup as an example in which we probe the eigenstates using the transmission of radiation, as well as the power spectrum of emitted radiation. Finally, we simulate numerically the pulsed direct excitation spectroscopy measurement of the two-excitation manifold performed in Ref.~\cite{Zanner2021}. All results for transmons are compared against the corresponding results for systems of qubits and harmonic oscillators, and the differences are discussed. The work is summarized in Sec.~\ref{sec:conclusions}

Details on the derivation of the collective master equation for a system of transmons inside a rectangular waveguide are given in App.~\ref{sec:master}. In App.~\ref{sec:nonherm} we discuss how the linear algebra of quantum mechanics has to be modified in order to describe non-Hermitian systems, which results in biorthogonal quantum mechanics~\cite{Brody_2013}. Finally, in App.~\ref{sec:numerics} we describe the methods used for numerically solving the dynamics encountered in this work.

\section{Waveguide QED for transmons}\label{sec:waveguideQED}
\subsection{Transmon array}
An array of~$L$ uncoupled transmons~\cite{Koch2007} is described accurately by the Bose--Hubbard Hamiltonian~\cite{Roushan17, Hacohen15, Ma2019, Fedorov2021},
\begin{equation}\label{eq:bose-hubbard}
   \frac{\hop_{\rm BH}}{\hbar} = \sum_{j=1}^L\omega_{j}\nop_{j}
    -\sum_{j=1}^L\frac{U_j}{2}\nop_j(\nop_j-1) 
\end{equation}
where~$\aop_{j}$ and~$\aop_j^\dag$ are the bosonic annihilation and creation operators of the site~$j$, with the commutation relation~$[\aop_j^{},\aop_k^\dag]=\delta_{jk}$, and~$\nop_j=\aop_{j}^\dag\aop_{j}$ is the corresponding number operator. Parameter $\omega_j$ is the transition frequency between the ground state and the first excited state of the~$j$th transmon, and~$U_j$ is the corresponding anharmonicity describing the on-site interactions. If the separation between transmon is sufficiently small, they couple capacitively to each other, which allows the hopping of excitations. This behavior can be included in Eq.~\eqref{eq:bose-hubbard} by adding a term~$\sum_{j\neq k} J^{}_{jk}\aop_{j}^\dag\aop^{}_{k}$, where~$J_{jk}$ is the hopping rate between sites~$j$ and~$k$.

For many-body dynamics the anharmonicity $U$ serves as a negative on-site interaction. Thus, the many-body interactions are attractive, in contrary to the repulsive model encountered in many atomic systems~\cite{Chen2016, Landig2016, Dogra2016, Baier2016, Bloch2008, Stoferle2004, Greiner2002, Landig2016}. As a single device, a transmon can be considered an intermediate between a harmonic oscillator and a qubit, and the strength of anharmonicity compared to the coupling strengths determines how close a transmon is to either of the limiting cases. The weaker the~anharmonicity~$U$, the more harmonic the system is, and in the opposite limit the transmon is more qubit like.

Because the excitations are bosons, occupations per site are not limited, and the lowest energy of the transmon system is obtained if all the excitations occupy the same site. However, only roughly ten lowest levels of a single transmon are bound~\cite{Leghtas1, Leghtas2, Pietikainen2019}, and they can be modeled using Eq.~\eqref{eq:bose-hubbard}. The Bose--Hubbard model thus breaks down if the total number of quanta in the system exceeds $10$, but in this work we do not consider such large fillings.

\subsection{Coherent interaction with electromagnetic field}
 To respect the bosonic multi-level nature of transmons, we derive the effective waveguide theory for a system of $L$ artificial atoms, each with $\lev$ arbitrarily spaced energy levels.  Neglecting the possible direct couplings between the atoms, the Hamiltonian describing the system is 
\begin{equation}
    \hop_{\rm sys} = \sum_{j=1}^L\sum_{m=0}^{d-1}E_{mj}\spop^{mj}\smop^{mj},
\end{equation}
where the operator~$\smop^{mj}=\ket{m_j}\bra{(m+1)_j}$ is the lowering operator for the~$(m+1)$st state of the~$j$th site, and~$E_{mj}$ is the energy of the~$m$th state of~$j$th site.

Propagating modes of the electromagnetic field inside the rectangular waveguide are described by the Hamiltonian~\cite{lalumiere2013, Pichler2015}
\begin{equation}\label{eq:H_atoms}
    \hop_{\rm F} =  \int_{-\infty}^\infty  dk_z \hbar\omega({k_z})\aop_{k_z}^\dag\aop^{}_{k_z},
\end{equation}
where~$k_z$ is the wavenumber in the~$z$ direction parallel to the waveguide (we assume that the waveguide is infinitely long), and $\aop_{k_z}$ is the corresponding annihilation operator for a radiation mode. The frequency of a mode is given by the non-linear dispersion relation
\begin{equation}\label{eq:H_field}
  \omega (k_z)    = \sqrt{c^2k_z^2 + \Omega_{\perp}^2},
\end{equation}
where~$c$ is the speed of light and~$\Omega_{\perp}$ is the cutoff frequency arising from the perpendicular dimensions of the rectangular waveguide, as discussed in App.~\ref{sec:EM_WG}. The situation is different whether the transition frequencies of the transmons are above or below the cutoff frequency, since no propagating electromagnetic modes exist with frequencies less than the cutoff frequency $\Omega_{\perp}$.

Next we introduce the coherent interaction of the transmon array and the electromagnetic environment in the waveguide, as schematized in Fig.~\ref{fig:circuit}(a).~The coupling between the field and the transmons is described by the Hamiltonian~\cite{lalumiere2013, Pichler2015}
\begin{equation}\label{eq:H_atom_field}
   \hop_{\rm I}
    = \sum_{mj}\hbar g_j\sqrt{m+1}\left(\xiop^{}_j+\xiop_j^\dag\right)
    \sxop^{mj},
\end{equation}
where~$g_j$ is a unitless coupling strength between the~$j$th transmon and the field. The operator~$\xiop_j$ is related to the electric field at the location~$z_j$ of the atom:
\begin{equation}
    \xiop_j = -i\int_{-\infty}^\infty dk_z\sqrt{\omega({k_z})}
    e^{ik_z z_j}\aop_{k_z}.
\end{equation}
From these, by following Ref.~\cite{lalumiere2013} (see App.~\ref{sec:master} for details), we trace out the electromagnetic environment and recover the master equation for the density operator of the transmons only~\cite{lalumiere2013, Mirhosseini2019, Gu2017, Pichler2015},
\begin{align}
  \frac{d\dens}{dt} =-i& \left[\frac{\hop_{\rm sys}}{\hbar}+\sum_{mj,nk} J_{mj,nk}\spop^{nk}\smop^{mj} ,\dens \right] \label{eq:gen_WQED}\\ 
                    +&\sum_{mj,nk}\gamma_{mj,nk}\left(\smop^{mj}\dens\spop^{nk}-\frac{1}{2}\left\{\spop^{nk}\smop^{mj},\dens\right\}\right).\notag 
\end{align}
Here~$J_{mj,nk}$ are the exchange interactions mediated by the waveguide, and~$\gamma_{mj,nk}$ are the collective damping rates arising from the interaction between the transmons and the radiation field.

\subsection{Long-range interaction and collective dissipation rates above and below the cutoff frequency}
Let us first consider the case with all the transition frequencies above the cutoff frequency, $\omega_{mj}>\Omega_\perp$. The radiation field of the rectangular waveguide can mediate long-range collective dissipation and interaction, similarly as with one-dimensional transmission lines~\cite{lalumiere2013}. The rates depend on the site separation and system frequencies as
\begin{align}
    \gamma_{mj,nk} =&\sqrt{\frac{\gamma_j\gamma_k}{\omega_j\omega_k}}\sqrt{(m+1)(n+1)} \notag \\
    &\times\sin\left(\frac{\pi x_j}{a}\right)
    \sin\left(\frac{\pi x_k}{a}\right)
    \left(\chi_{mjk}+\chi_{nkj}^*\right),\\
    J_{mj,nk} =&-\frac{i}{2}\sqrt{\frac{\gamma_j\gamma_k}{\omega_j\omega_k}}\sqrt{(m+1)(n+1)} \notag\\
    &\times\sin\left(\frac{\pi x_j}{a}\right)
    \sin\left(\frac{\pi x_k}{a}\right)
    \left(\chi_{mjk}-\chi_{nkj}^*\right),
\end{align}
where for the $j$th emitter, we have the single site decay rate~$\gamma_j=4\pi\omega_jg_j^2$, a representative frequency~$\omega_j$, the coordinate~$x_j$ along the width~$a$ of the waveguide, perpendicular to the radiation propagation, see Fig~\ref{fig:circuit}(a). The collective dissipation and interaction rates are defined through oscillatory coefficients  
\begin{equation}\label{eq:chi}
    \chi_{mjk} = \frac{\omega_{mj}^2}
    {\sqrt{\omega_{mj}^2-\Omega_\perp^2}}
    e^{it_{jk}\sqrt{\omega_{mj}^2-\Omega_\perp^2}},
  \end{equation}
where $t_{jk} = |z_j-z_k|/c$ is the propagation time in empty space between locations of the sites~$j$ and~$k$, and~$\omega_{mj}=(E_{m+1,j}-E_{mj})/\hbar$ is the transition frequency between~$(m+1)$st and~$m$th eigenlevels of site~$j$, which for the transmon depends on the anharmonicity~$U$ as~$\omega_{mj} = \omega_j - mU_{j}$. 

Assuming that the emitters are located at the center line, $x_j=a/2$, their transition frequencies are homogeneous and weakly anharmonic $\omega_{mj}=\omega_{nk}\approx\omega_0$, and they are far from the cutoff frequency $\omega_{mj}\approx \sqrt{\omega_{mj}^2-\Omega_\perp^2}$, we obtain the expressions~\cite{lalumiere2013, Mirhosseini2019}
\begin{align}
    \gamma_{mj,nk} &= \sqrt{\gamma_j \gamma_k}\sqrt{(m+1)(n+1)} \cos (\omega_0 t_{jk}),\label{eq:gamma_mjnk}\\  
    J_{mj,nk} &= \frac{\sqrt{\gamma_j\gamma_k}}{2}\sqrt{(m+1)(n+1)} \sin (\omega_0 t_{jk}).\label{eq:J_mjnk}
\end{align}
The dissipation rate and interaction strength are oscillatory functions in terms of the phase difference between the sites, which can be controlled either via the frequency of the emitters or their separation, best seen by writing $\omega_0 t_{jk}=2\pi|z_j-z_k|/\lambda_0$ in terms of the wavelength $\lambda_0$. Thus, if the site separation is an integer multiple of half of the wavelength~$\lambda_0$, the correlated decay obtains its maximal value, whereas the exchange interaction is at minimum. The situation is reversed if the site separation is an odd multiple of quarter of the wavelength, in which case the correlated decay is at minimum and exchange interaction at maximum. Even though the correlated decay vanishes in this case, each site still decays individually. For evenly spaced spectrum the coefficients would vanish at minimum, but for anharmonic transitions there can be weak exchange interaction also with maximal correlated decay, and vice versa. In this paper we focus on the situation with maximal correlated decay.

Additionally, when the waveguide is driven from left, the array becomes also effectively driven, described by the Hamiltonian (see App.~\ref{sec:master_equation_app})
\begin{equation}\label{eq:wg_input}
  \hat{H}_{\rm d}(t)
  =-\sum_{mj}\sqrt{\frac{2\hbar \gamma_{mj,mj}}{\omega_{mj}}}\sqrt{P}\sin\left[\omega_{\rm d}(t+t_j)\right]\hat\sigma_x^{mj},
\end{equation}
where~$P$ is the power of the radiation,~$\omega_{\rm d}$ is the frequency of the input, and~$t_j = z_j/c$ is the time it takes for a photon to propagate to site~$j$. For the first site one can set~$t_1=0$, since the positions here only determine the phase at each site.

Below the cutoff frequency when $\omega_{mj}<\Omega_\perp$, we find the dissipation and interaction rates similarly,
\begin{align}
    \gamma^\perp_{mj,nk} =-&i\sqrt{\frac{\gamma_j\gamma_k}{\omega_j\omega_k}} \sqrt{(m+1)(n+1)} \notag\\
    &\times  \sin\left(\frac{\pi x_j}{a}\right)
    \sin\left(\frac{\pi x_k}{a}\right)
    \left(\zeta_{mjk}-\zeta_{nkj}\right),\\
    J^\perp_{mj,nk} =-&\frac{1}{2}\sqrt{\frac{\gamma_j\gamma_k}{\omega_j\omega_k}} \sqrt{(m+1)(n+1)} \notag \\
    & \times  \sin\left(\frac{\pi x_j}{a}\right)
    \sin\left(\frac{\pi x_k}{a}\right)
    \left(\zeta_{mjk}+\zeta_{nkj}\right),
\end{align}
which decay exponentially with the site separation,
\begin{equation}
    \zeta_{mjk} = \frac{\omega_{mj}^2}
    {\sqrt{\omega_{mj}^2-\Omega_\perp^2}}
    e^{-t_{jk}\sqrt{\Omega_\perp^2-\omega_{mj}^2}}.
\end{equation}
The dissipation  and interaction rates reduce with the assumptions $x_j=a/2$ and $\omega_{mj}=\omega_{nk}\approx \omega_0$ to
\begin{align}
  \gamma^\perp_{mj,nk}&=0,\\
   J^\perp_{mj,nk} &=-\sqrt{\frac{\omega^2_0\gamma_j\gamma_k}{\Omega_\perp^2-\omega^2_0}}
   \frac{\sqrt{(m+1)(n+1)}}{2}
   e^{-t_{jk}\sqrt{\Omega_\perp^2-\omega_0^2}}.
\end{align}
Below the cutoff frequency, radiation does not propagate and energy cannot leak out from the array via the waveguide, nor can the array be driven through the waveguide. Thus, dissipation is not possible, seen by identically zero dissipation rates. However, the exchange interaction is not identically zero, but its strength decays exponentially as a function of the site separation. In other words, nearby emitters can still exchange excitations via the waveguide, in addition to the capacitive coupling. 

For the rest of this paper we assume that all the transmons are sufficiently far above the cutoff, so that we can set~$\Omega_\perp=0$ in Eq.~\eqref{eq:chi}, recovering the dispersionless propagation.

\section{Collective bosonic many-body effects}\label{sec:collective}
In this section we study how an array of weakly anharmonic oscillators, such as transmons, behaves under the influence of the collective electromagnetic environment of a waveguide, and compare the results to the widely studied qubit case, especially the Dicke model~\cite{Bowden1979,Gilmore1978,Domokos2002,Klinder2015,Emary2003,Nagy2010,Strack2011,Bhaseen2012,Bastidas2012,Kulkarni2013,Kloc2018,Kirton2019,Rosatto2020,Lemberger2021,Mlynek2014,Higgins2014} and to the case of an array of harmonic oscillators.

We start by noting that the master equation~\eqref{eq:gen_WQED} can be reformulated as 
\begin{equation}
    \frac{d\dens}{dt} =-\frac{i}{\hbar}\left(\hop_{\rm eff}\dens-\dens \hop_{\rm eff}^\dag\right)+\sum_{mj,nk}\gamma_{mj,nk}\smop^{mj}\dens\spop^{nk}, \label{eq:gen_WQED_2}
\end{equation}
where the non-Hermitian effective Hamiltonian~\cite{Mirhosseini2019} is
\begin{equation}
    \hop_{\rm eff}=\hop_{\rm sys}+\hbar\sum_{mj,nk} \left(J_{mj,nk}- \frac{i\gamma_{mj,nk}}{2}\right)\spop^{nk}\smop^{mj}. \label{eq:effective_hamiltonian}
\end{equation}
The dynamical behavior of the system can then be understood with the quantum trajectory description~\cite{Molmer93, Daley2014}. The latter part of the master equation~\eqref{eq:gen_WQED_2} describes quantum jumps, i.e., the collective decay events in which the waveguide radiation field transports energy from the array. The non-Hermitian effective Hamiltonian describes the non-unitary time evolution between the quantum jumps, and it has complex-valued eigenvalues $\hat H_{\rm eff}\ket{\alpha}=\lambda_\alpha \ket{\alpha}$ of the form 
\begin{equation}\label{eq:complex_eigenvalues}
    \lambda_\alpha = E_\alpha - i\hbar\frac{\Gamma_\alpha}{2},
\end{equation}
where~$E_\alpha$ is interpreted as the energy and~$\Gamma_\alpha$ as the total decay rate of the state~$\ket{\alpha}$. Non-Hermitian quantum mechanics is discussed in App.~\ref{sec:nonherm} more thoroughly. Similarly as with unitary quantum dynamics, the eigenvalues of the effective Hamiltonian determine the behavior of the dissipative quantum system~\cite{Molmer2019, Molmer2020, Celardo2013, Celardo2013_3D}. Specifically, considering a short time interval $d t$, the decay rate specifies the decay probability  $P_\alpha(d t)=\Gamma_{\alpha} d t$, and the quantum jump term of the master equation determines the details of the decay process. 

For better analytical understanding, we simplify the master equation~\eqref{eq:gen_WQED_2}. First, we assume that all the sites are identical, so that they have the same frequencies, and they all couple to the waveguide with $\gamma=\gamma_j$. We also assume that the coefficients~$\gamma_{mj,nk}$ and~$J_{mj,nk}$ are equal for all~$m$ and~$n$, i.e., for bosonic systems we assume that the eigenlevels of the sites are evenly spaced. Then the effective Hamiltonian is expressed for bosonic systems in terms of the annihilation operators ~$\aop_j = \sum_{m=0}^\infty \sqrt{m+1}\smop^{mj}$ and for the qubit systems with the corresponding $\sigma_-^j=\smop^{0j}=\ket{0_j}\bra{1_j}$. Finally, we assume that the sites are spaced by a distance which is an integer $n$ multiple of the wavelength, $|z_j-z_k|=n\lambda_0$ (see Fig.~\ref{fig:circuit}), so that each site has the same phase, implying $J_{j,k}=0$ and $\gamma_{j,k}=\gamma$ for all~$j$ and~$k$, including the possibility that the sites are at the same location. This assumption is lifted later in Sec.~\ref{sec:coupled_system}.

\begin{figure}[t]
    \centering
    \includegraphics[width=1.0\linewidth]
    {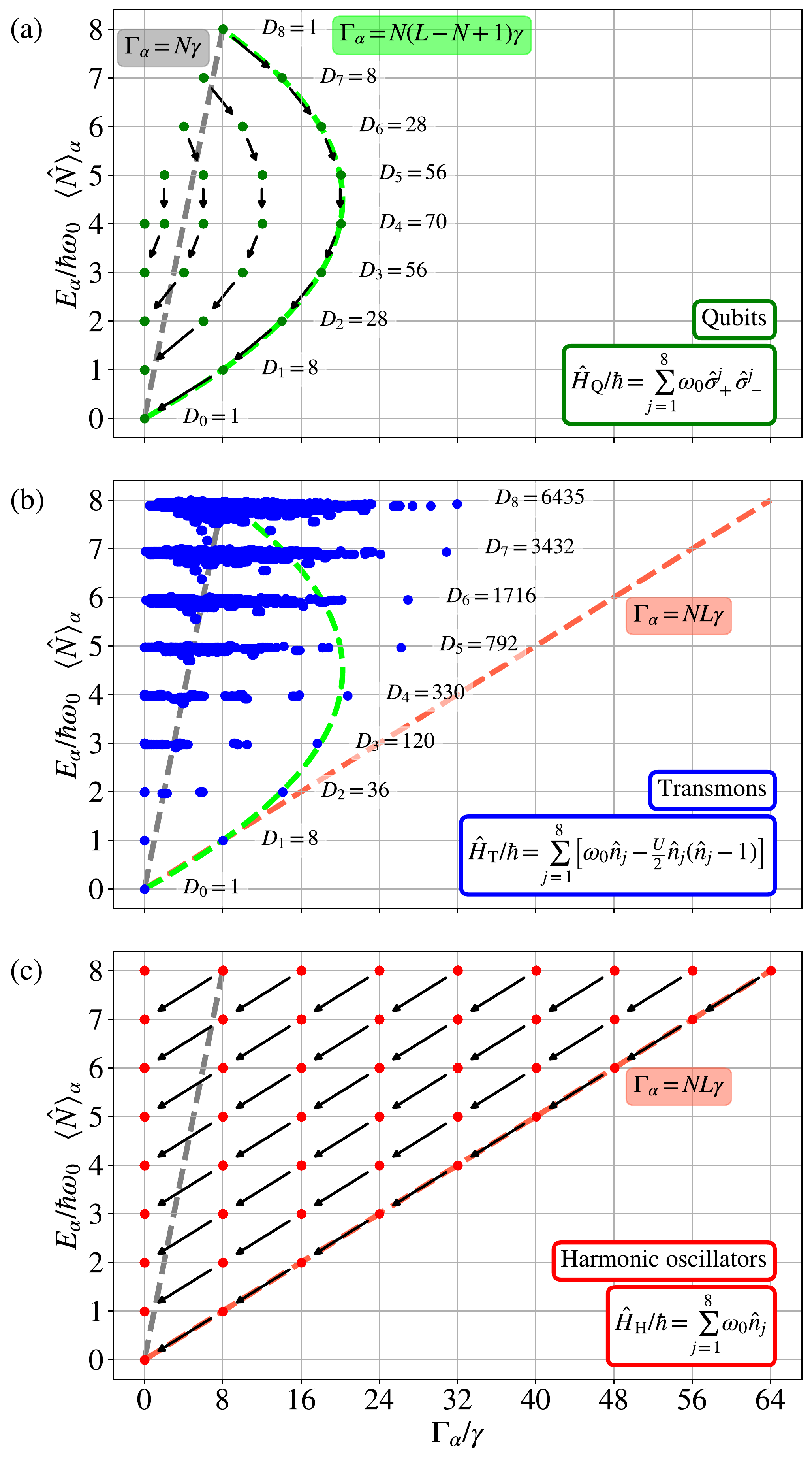}
    \caption{Complex eigenvalues of the effective non-Hermitian Hamiltonian as a function of the energy $E_{\alpha}$ and the decay rate $\Gamma_{\alpha}$ for $L=8$ uncoupled (a) qubits, (b) transmons with~$U=8.72\gamma$, and (c) harmonic oscillators.  For qubits the decay rates of the brightest states depends quadratically on the occupation~$N$ (dashed green curve), for the harmonic oscillators we see linear scaling (dashed red line). For transmons the decay rates scale similarly to qubits up to half filling, after which they continue to increase. For comparison also the decay rates without correlated effects are shown as dashed grey lines. Hilbert space dimensions for each excitation manifold are also shown, the dimensions are the same in (b) and (c).}
    \label{fig:levels_qubit_ho_transmon}
\end{figure}

To summarize, here we contrast the qubit and bosonic models through the effective non-Hermitian Hamiltonians
\begin{align}
 \hop^{\rm Q}_{\rm eff} &= \hop_{\rm Q} - i\hbar  \frac{\gamma}{2}\sum_{j,k}\sigma_+^k\sigma_-^j, \label{eq:effective_hamiltonian1_qub}\\
  \hop^{\rm B}_{\rm eff}&= \hop_{\rm H/T} - i\hbar \frac{\gamma}{2}\sum_{j,k}\aop_k^\dag\aop^{}_j. \label{eq:effective_hamiltonian1_bos}
\end{align}
In Fig.~\ref{fig:levels_qubit_ho_transmon} we plot the complex eigenvalues of the non-Hermitian Hamiltonians on the $\Gamma_\alpha-E_\alpha$~-plane for three different cases, where uncoupled emitters inside the waveguide are either (a)~qubits, (b)~transmons, or (c)~harmonic oscillators. Without correlated decay ($\gamma_{j,k}=0,\ j\neq k$), the effective Hamiltonian would be $\hat H^{\rm B}_{\rm eff}=\hat H_{\rm B}-i \hbar \frac{\gamma}{2}\sum_{j}\hat a^\dag_j \hat a^{}_j$ for bosonic systems, and~$\hat H^{\rm Q}_{\rm eff}=\hat H_{\rm Q}-i \hbar \frac{\gamma}{2}\sum_{j}\spop^j \smop^j$ for qubits. In such systems one observes a linear scaling as a function of total occupation~$N$ in the decay rates $\Gamma$, so that all the eigenvalues lie on the line $\Gamma=N\gamma$, where $N=\braket{\sum_{j}\aop_{j}^\dag \aop_{j}^{}}$ for bosonic systems and $N=\braket{\sum_j(\hat I +\hat \sigma_z^j)/2}$ for qubits, see grey dashed diagonal line in Fig.~\ref{fig:levels_qubit_ho_transmon}. Correlated decay causes some of the states to decay faster (superradiance) or slower (subradiance) than $\gamma N$.

\subsection{Collective decay in an array of qubits}\label{sec:cd_qub}
The system of~$L$ identical qubits in a waveguide is represented by the Hamiltonian
\begin{equation}\label{eq:Q_uncoupled}
    \frac{\hop^{\rm Q}_{\rm eff}}{\hbar} = 
    \sum_{j=1}^L\frac{\omega_0}{2}\left(\hat{I}+\szop^j\right) 
    - i\frac{\gamma}{2}\sum_{j=1}^L\sum_{k=1}^L\spop^j\smop^k,
\end{equation}
where~$\szop^j$ is the Pauli~$z$ matrix of the~$j$th qubit and~$\smop^j$ is the associated lowering operator. The eigenstates are~$\ket{s,m_z}$ where~$m_z$ is an eigenvalue of the~$z$ component of the total spin~$\hat S_z=\sum_{j=1}^L\szop^j$ and~$s$ is related to the eigenvalue of the length of the total spin~$\hat S^2=\hat S^2_x+\hat S^2_y+\hat S^2_z$. The possible values are~$s=L, L-2, \ldots,0$ and $m=s,s-2, \ldots, -s$. We also define~$\hat S_-=\sum_{s=1}^L\smop^j$ as the total lowering operator. With these total spin operators, the Hamiltonian~\eqref{eq:Q_uncoupled} can be written as
\begin{equation}\label{eq:Q_uncoupled2}
    \frac{\hop^{\rm Q}_{\rm eff}}{\hbar} = \frac{\omega_0}{2}(\hat S_z+\hat I L) 
    - i\frac{\gamma}{2}\hat S_+\hat S_-,
\end{equation}
from which we can see that the state~$\ket{s,m_z}$ has an energy~$E_{sm_z}=\hbar\omega_0(m_z+L)/2=\hbar N\omega_0$ and decay rate~$\Gamma_{sm_z}=\gamma (s+m_z)(s-m_z+2)/4$. 
The brightest state of the~$N$-excitation manifold has a decay rate~$N(L-N+1)\gamma$, see dashed green curve in Fig.~\ref{fig:levels_qubit_ho_transmon}(a). These states have the maximal total spin~$\ket{s=L,m_z}$. The dark states with $\Gamma_\alpha=0$ are the states~$\ket{s,m_z=-s}$ with the lowest possible value for the total $\hat S_z$, and they can only exist up to half-filling~$N=L/2$~\cite{Poddubny2019}. 

The decay of the collective system is caused by the operator~$\hat S_-$, denoted by black arrows in Fig.~\ref{fig:levels_qubit_ho_transmon}(a). The cascaded collective decay forms decay manifolds: $\ket{s,m_z}\to\ket{s,m_z-2}\to\ldots$. 
For example, there is only one state with~$N=L$ excitations, $\ket{s=L,m_z=L}$. If the system is initially in this state, it decays with the rate~$N\gamma$ to the state $\ket{s=L,m_z=L-2}$ in $N=L-1$ manifold which has a larger decay rate. As we go down the decay ladder, the decay rate first increases, reaches its maximum at half-filling, and then starts to decrease again. This is observed in Fig.~\ref{fig:levels_qubit_ho_transmon}(a) as the parabolic scaling of the decay rates as a function of the number of excitations.

\subsection{Collective decay in an array of harmonic oscillators}\label{sec:cd_harm}
The effective Hamiltonian of an array of harmonic oscillators reads
\begin{equation}\label{eq:HO_uncoupled}
    \frac{\hop^{\rm H}_{\rm eff}}{\hbar} = \sum_{j=1}^L\omega_0\nop_j - i\frac{\gamma}{2} \sum_{j=1}^L\sum_{k=1}^L\aop_j^\dag\aop^{}_k.
\end{equation}
In solving and analyzing the eigenstates of the system we utilize the fact that the non-Hermitian Hamiltonian of Eq.~\eqref{eq:HO_uncoupled} conserves the total boson number, that is, it commutes with the total occupation operator~$\ntot$,
\begin{equation}
    \left[\hop_{\rm eff}^{\rm H},\ntot \right]=0,\quad\ntot=\sum_{j=1}^L\nop_j.
\end{equation}
Because of this, the effective Hamiltonian is block-diagonal, where each block is characterized by a total number of excitations,~$\braket{\ntot} = N$. The number of states in an $N$-excitation manifold is 
\begin{equation}
    D_{N,L}=\binom{N+L-1}{N}=\frac{(N+L-1)!}{(L-1)!N!} \label{eq:dnl}
\end{equation}
We now diagonalize the effective Hamiltonian~\eqref{eq:HO_uncoupled} with collective bosonic operators
\begin{equation}\label{eq:collective_ops}
    \cop_k = \frac{1}{\sqrt{L}}\sum_{j=1}^{L}
    \exp\left(\frac{2\pi i}{L}jk\right)\aop_{j},
\end{equation}
where $k=1,2,\dots,L$ and $[ \hat c^{}_k,\hat c_{k'}^\dag ]=\delta_{k,{k'}}$. The result is a set of $L$ uncoupled harmonic oscillators
\begin{equation}
    \frac{\hop^{\rm H}_{\rm eff}}{\hbar} =\sum_{k=1}^L\omega_0\cop_k^\dag\cop^{}_k 
    -i \frac{L\gamma}{2} \cop_L^\dag\cop^{}_L  \label{eq:effective_HO}
\end{equation}
Only the mode corresponding to $\cop_L$ decays at rate $L \gamma$, the other modes are dark. The collective eigenstates of the non-Hermitian effective Hamiltonian~\eqref{eq:effective_HO} are 
\begin{equation}\label{eq:HO_collective_states}
    \ket{m_1,m_2,\dots,m_{L}}\\
     = \frac{\left(\cop_1^\dag\right)^{m_1}\left(\cop_2^\dag\right)^{m_2}\cdots\left(\cop_L^\dag\right)^{m_L}}{\sqrt{m_1!m_2!\cdots m_L!}}\ket{G},
\end{equation}
where $\ket{G}$ is the ground state. Each quantum generated by the operator~$\cop_L^\dag$ increases the decay rate of the corresponding state by~$L\gamma$. The brightest superradiant state with~$N$ excitations is
\begin{equation}
    \ket{\rm{SR}_{(N)}^{\rm H}}=\ket{0,0,\dots,N}=\frac{\left(\cop_L^\dag\right)^N}{\sqrt{N!}}\ket{G}.
\end{equation}
There exists only one such a state for a given~$N$, see the rightmost filled circles in Fig.~\ref{fig:levels_qubit_ho_transmon}(c). Considering these states as a function of the total boson number $N$ and the total number of sites $L$, the decay rates of the brightest states scale linearly as $\Gamma_{\rm max}^{\rm H}=NL\gamma$, see the red diagonal dashed line in Fig.~\ref{fig:levels_qubit_ho_transmon}(c). It is~$L$ times larger than without the correlated effects~\cite{delanty2011superradiance} (grey dashed line) and much larger than the collective decay rate of the qubits $\Gamma_{\rm max}^{\rm Q}=N(L-N+1)\gamma$ for $N>1$, see the green dashed curve in Fig.~\ref{fig:levels_qubit_ho_transmon}(a). Thus, for bosonic systems the behavior of the collective decay is fundamentally different compared to the case of qubits. 

The difference between the bosonic and qubit superradiance can be understood through local bosonic multi-occupancy, which results in bosonic enhancement of decay rates. For example, for $L$ oscillators and $N=2$ excitations, the most superradiant bosonic state  is 
\begin{equation}
    \ket{\rm{SR}_{(2)}^{\rm H}}  =\frac{1}{L}\bigg( \sqrt{2}\sum_{\ell=1}^L\sum^{l-1}_{\ell'=1}\ket{n_\ell=1,n_{\ell'}=1} + \sum_{\ell=1}^L\ket{n_\ell=2} \bigg),
\end{equation}
where $\ket{n_\ell=n}=\left(\aop_{\ell}^\dag\right)^n\ket{G}/\sqrt{n!}$, whereas the corresponding most superradiant qubit state is otherwise identical but misses the doubly occupied states,
\begin{equation}
    \ket{\rm{SR}_{(2)}^{\rm Q}}  = \sqrt{\frac{2}{L(L-1)}} \sum_{\ell=1}^L\sum^{\ell-1}_{\ell'=1}\ket{n_\ell=1,n_{\ell'}=1}_{\rm Q},
\end{equation}
where $\ket{n_\ell=1}_{\rm Q}=\hat{\sigma}_+^\ell\ket{G}$.
The dark states are all the states of Eq.~\eqref{eq:HO_collective_states} where $m_L=0$, meaning that no quanta are created by~$\cop_L^\dag$. They are multiply degenerate, and as opposed to qubits, there exists dark states in all excitation manifolds. This occurs because of the larger Hilbert space dimension of the bosonic system, which allows multiple occupations per site. For example at $N=L=2$, the bosonic dark state $\Gamma=0$ is
\begin{equation}
  \ket{\rm{SUB}^{\rm H}_{(2)}}=\frac{1}{2}\left(\ket{20}-\sqrt{2}\ket{11}+ \ket{02}\right).
\end{equation}
In the corresponding qubit case, the only state is $\ket{11}$ which is decaying with the rate $\Gamma=2\gamma$ as an unentangled state that cannot have any correlated effects, see Fig.~\ref{fig:circuit}(b). 

In general, it is possible to have states with a decay rate $mL\gamma$, where $0\le m \le N$. The number of such states is given by the formula
\begin{equation}\label{eq:number_of_decay}
    d_{m, N, L} = \frac{(N-m+L-2)!}{(N-m)!(L-2)!} = D_{N-m,L-1},
\end{equation}
where~$D_{m,L}$ is the number of bosonic states, defined in Eq.~\eqref{eq:dnl}. From Eq.~\eqref{eq:effective_HO} we see that only one collective operator,~$\cop_L$, causes jumps between different excitation manifolds,
  \begin{equation}
        \cop_L\ket{m_1,m_2,\dots,m_{L}}=\sqrt{m_L}\ket{m_1,m_2,\dots,m_{L}-1},
\end{equation}
i.e., the decay happens such that~$\cop_L$ removes one quantum from the state. This causes the diagonal decay ladders, as observed in Fig.~\ref{fig:levels_qubit_ho_transmon}(c).

\subsection{Collective decay in an array of transmons}\label{sec:cd_trans}
The effective Hamiltonian of an array of transmons is
\begin{equation}\label{eq:trans_uncoupled}
    \frac{\hop^{\rm T}_{\rm eff}}{\hbar} = \sum_{j=1}^L\left[\omega_0\nop_j -\frac{U}{2}\nop_j(\nop_j-1)\right]-i\frac{\gamma}{2} \sum_{j=1}^L\sum_{k=1}^L\aop_j^\dag\aop^{}_k.
\end{equation}
Transmons differ from harmonic oscillators through the weak anharmonicity~\cite{Koch2007} giving rise to the interaction term $-(\hbar U /2)\sum_j\nop_j(\nop_j-1)$ in the many-body setting. Now, the interaction term and the collective decay terms do not commute, 
\begin{equation}
    \left[\frac{U}{2}\sum_{j=1}^L\nop_j(\nop_j-1),\frac{\gamma}{2} \sum_{j=1}^L\sum_{k=1}^L\aop_j^\dag\aop^{}_k \right]\neq 0,
\end{equation}
which implies that the eigenstates of the non-Hermitian effective Hamiltonian are neither the eigenstates of the uncoupled transmon array nor the eigenstates of the collective decay term. 

Typically the interaction strength $U$ dominates over the collective decay strength $\gamma$, $U/\gamma \gtrsim 10$~\cite{Zanner2021}, which would suggest that the eigenstates could be solved by considering the collective decay as a perturbation. However, the situation is more complicated than that due to the high number of many-body Fock states that are degenerate with respect to the interaction term. For example, in the manifold of total $N$ excitations, states with all possible permutations for the state occupations similar to $\ket{N-2, 1, 1, 0,\ldots, 0}$ are degenerate. The situation is similar to the problem of solving the ground states of the Bose--Hubbard model with attractive interactions~\cite{Mansikkamaki2021} where two phases emerge, the delocalized superfluid or the localized W phase, depending on the strength of the hopping rate (which here corresponds to the collective decay strength) with respect to the interaction strength. Furthermore, we are interested on the full complex spectrum instead of just the ground states, rendering the problem even more challenging. Hence, in this section, we resort only on the numerical solution displayed in Fig.~\ref{fig:levels_qubit_ho_transmon}(b) of the array of $L$ transmons. In Sec.~\ref{sec:twopairs} we focus in more detail on the case of four transmons. 

The complex spectrum of a transmon array is somewhere in between that of qubits and harmonic oscillators. Unlike with qubits, we observe dark states also beyond half-filling, and the decay rates of the brightest states grow as a function of~$N$, although not as strongly as with harmonic oscillators (red dashed line). The high number of degeneracy observed with the harmonic oscillators is reduced due to the anharmonicity~$U$ of the transmons, which decreases the energies. The brightest states of a transmon array are in general at high energy in each excitation manifold. This means that the attractive many-body interaction affects them only slightly, meaning that most of the contribution to the brightest states comes from the Fock states where the excitations are spread out to the sites as evenly as possible.  For example, in the case of $N=L$, the most superradiant states are the superpositions of mainly the Fock states that are different permutations of $\ket{111\ldots 1}$ and $\ket{1021\ldots 1}$. The large bosonic many-body Hilbert space of transmon arrays thus allows the construction of states that are either much more subradiant or superradiant than in the corresponding qubit array case. 

On the other hand, the lowest energy states of each excitation manifold lie approximately at the line~$\Gamma_\alpha = N\gamma$, which gives the decay rate of the state without the correlated effects. The many-body interactions thus decrease the collective behavior in bosonic systems.

\section{Interplay between local and global collective states}\label{sec:interplay}
Above we studied the case where the transmons are spaced an integer multiple of the wavelength apart, $|z_j-z_k|=n\lambda_0$, such that the waveguide mediated exchange interactions $J_{mj,nk}$ vanish and the emitters are only connected through the collective decay terms. Another natural limit would be the case where the transmons are very close to each other such that the separations between any two sites is approximately zero, $|z_j-z_k|\approx 0$. The form of the collective decay terms is the same as with the integer wavelength case, but one should additionally take into account also the direct capacitive coupling~$J$ in Eq.~\eqref{eq:bose-hubbard} between the transmons, which is always present if the sites are sufficiently close to each other~\cite{Dalmonte2015}. For simplicity in analytical calculations, if such an additional term is present, one would like to have it such that it commutes with the collective decay term of the non-Hermitian effective Hamiltonian. Such couplings include equally strong all-to-all coupling (the collective decay is an all-to-all coupling) or a ring of transmons with each site coupled to its nearest neighbors. However, such systems can be difficult to realize in practice for a large number of sites. 

\begin{figure}[t]
    \centering
    \includegraphics[width=1.0\linewidth]{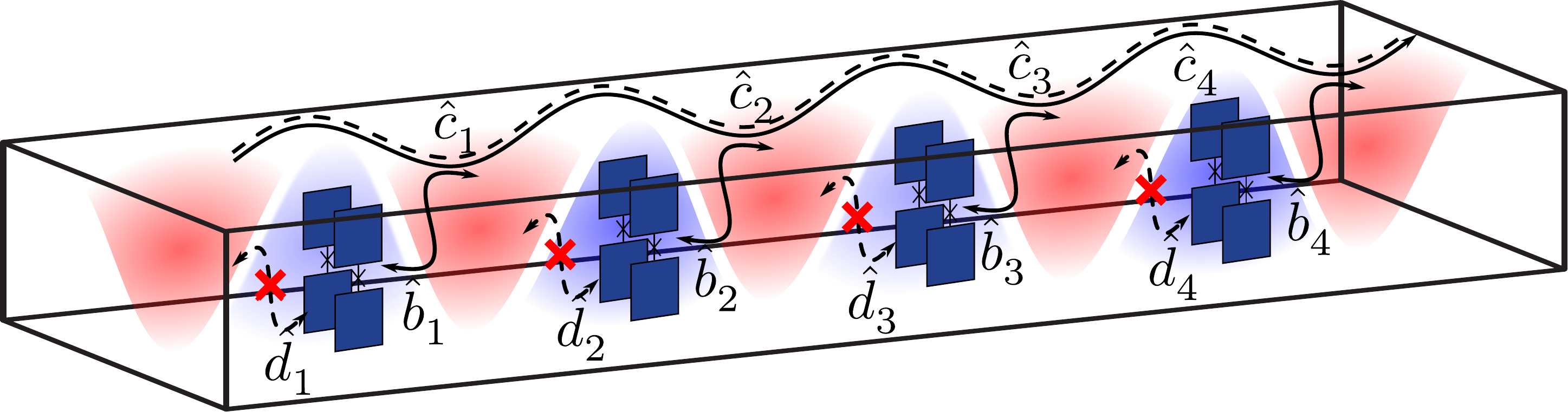}
    \caption{Schematic of an array of transmon pairs. In the one-excitation manifold each pair hosts one local dark state and one local bright state created by the operators~$\dop_j^\dag$ and~$\bop_j^\dag$, see Eq.~\eqref{eq:local_modes}. The local bright states further interact via the waveguide, resulting in three collective dark states created by the operators~$\cop_{1,2,3}^\dag$, and one collective bright state created by~$\cop_4^\dag$, see Eq.~\eqref{eq:collective_modes}.}
    \label{fig:multiple_pairs}
\end{figure}

One further possibility is to consider an array made of transmon pairs. The two closely located transmons form a capacitively coupled pair, and several of these pairs are evenly spaced along the waveguide to form an array, see Fig.~\ref{fig:multiple_pairs}. The motivation to study such a construction is its intriguing internal structure, where each transmon pair hosts local bright and dark states, and only the local bright states contribute in the formation of array-wide collective dark and bright states.~By having a side port control on the local transmons, as demonstrated in Ref.~\cite{Zanner2021}, one can imagine a scenario where quantum information stored on the local dark states is first converted to the local bright states.~The local bright and dark states are separated in energy and state symmetry, providing means for state specific addressing. Then the global dark states, formed from the local bright states, form a quantum bus to communicate between different transmon pairs. Further, the separation of local and global states in energy opens new possibilities for the implementation of quantum simulations.

\subsection{An array of transmon pairs}\label{sec:coupled_system}
 Without the interaction~$U$, the effective Hamiltonian for~$L$ transmon pairs reads 
    \begin{align}
         \frac{\hop^{\rm pairs}_{\rm eff}}{\hbar} &= \sum_{j=1}^{L} \Big[\omega_0\left(\nop_{j1}+\nop_{j2}\right)
        + J\left(\aop_{1j}^\dag \aop^{}_{2j}+\rm{h.c.}\right)\Big] \notag \\
        &-i\frac{\gamma}{2}\sum_{p,l=1}^2\sum_{k,j=1}^Le^{i\omega_0 t_{jk}}\aop_{pk}^\dag\aop^{}_{lj}, \label{eq:ham_multiple_pairs}    
    \end{align}
where~$L$ is now the number of pairs, and the indices~$j$ and~$k$ refers to the pair. We assume that the sites forming a pair are located at the same position in the waveguide, so that~$t_{jj} = 0$, and we take the separation of the pairs to be of the order of the wavelength corresponding to the frequency of the transmons, so that the pairs do not couple capacitively to each other, but interact only through the waveguide. Within a pair, the diagonalized local operators are 
\begin{align}
    \bop_j & =\frac{1}{\sqrt{2}}\left(\aop_{1j}+\aop_{2j}\right), & \dop_j &= \frac{1}{\sqrt{2}}\left(\aop_{1j}-\aop_{2j}\right).\label{eq:local_modes}
\end{align}
 In terms of these, the Hamiltonian of $L$ pairs becomes
\begin{align}
    \frac{\hop^{\rm pairs}_{\rm eff}}{\hbar} &= \sum_{j=1}^{L}\left[\left(\omega_0+J\right)\bop_j^\dag\bop^{}_j + \left(\omega_0-J\right)\dop_j^\dag\dop^{}_j\right] \notag \\
        &-i\gamma\sum_{j=1}^L\sum_{k=1}^Le^{i\omega_0t_{jk}}\bop_k^\dag\bop^{}_j,
\end{align}
where the operators~$\bop_j^\dag$ and~$\cop_j^\dag$ create an excitation on the local bright and dark modes of the~$j$th pair, respectively. The local modes are split by energy~$2\hbar J$, so that the bright states are higher in energy. Due to their nature, the dark states do not interact via the waveguide. The local bright states, on the other hand, combine to form system-wide collective states. 

Assuming that $\omega_0t_{jk} = 2\pi|j-k|$, so that the phase difference is the same for all pairs, we can write the collective operators as a Fourier series,
\begin{equation}\label{eq:collective_modes}
    \cop_k = \frac{1}{\sqrt{L}}\sum_{j=1}^{L}\exp\left(\frac{2\pi i}{L}jk\right)\bop_j,
\end{equation}
similarly as in Eq.~\eqref{eq:collective_ops}, so that the Hamiltonian becomes
\begin{align}
       \frac{\hop^{\rm pairs}_{\rm eff}}{\hbar} &=   \left(\omega_0+J-iL\gamma\right)\cop_L^\dag\cop^{}_L \notag  \\
       &+\sum_{j=1}^{L-1}\left(\omega_0+J\right)\cop_j^\dag\cop^{}_j + \sum_{j=1}^{L} \left(\omega_0-J\right)\dop_j^\dag\dop^{}_j.\label{eq:pairs_f}
\end{align}
We thus find one global bright mode~$\cop_L$ with decay rate~$2L\gamma$,~$L-1$ global dark modes $\cop_{1,2,\ldots, L-1}$, and $L$ local dark modes $\dop_j$. The complex spectrum of the Hamiltonian~\eqref{eq:pairs_f} is similar to that of the harmonic oscillators in Fig.~\ref{fig:levels_qubit_ho_transmon}(c) with the exception that now the local and global modes are split in energy, reducing the degeneracy. 

The interaction term $-(\hbar U /2)\sum_p\sum_j \nop_{jp} (\nop_{jp}-1)$ will give similar effects as for the array of wavelength spaced transmons, discussed in Sec.~\ref{sec:cd_trans}. Especially, it couples the local and global modes, which we will next elaborate in detail in the case of two transmon pairs. 

\subsection{Two pairs of transmons} \label{sec:twopairs}
\begin{table}[t]
    \centering
    \caption{Parameters of the transmon array and the waveguide, their respective symbols and values used in numerical calculations. The values are chosen close to the ones measured in Ref.~\onlinecite{Zanner2021}.}
    \begin{tabular}{lll}
        \hline
        Parameter & Symbol & Value\\
        \hline
        Transmon frequency & $\omega_0/2\pi$ & \SI{7.28}{\giga\hertz}\\
        Anharmonicity & $U/2\pi$ & \SI{218}{\mega\hertz}\\
        Capacitive coupling strength & $J/2\pi$ & \SI{45}{\mega\hertz}\\[0.1cm]
        
        Waveguide coupling strength & $\gamma/2\pi$ & \SI{25}{\mega\hertz}\\
        Waveguide cutoff frequency & $\Omega_\perp/2\pi$ & \SI{6.55}{\giga\hertz}\\[0.1cm]
        Bulk dissipation rate & $\kappa/2\pi$ & \SI{15}{\kilo\hertz}\\
        \hline
    \end{tabular}
    \label{tab:exp_params}
\end{table}

\begin{figure}[t]
    \centering
    \includegraphics[width=1.0\linewidth]{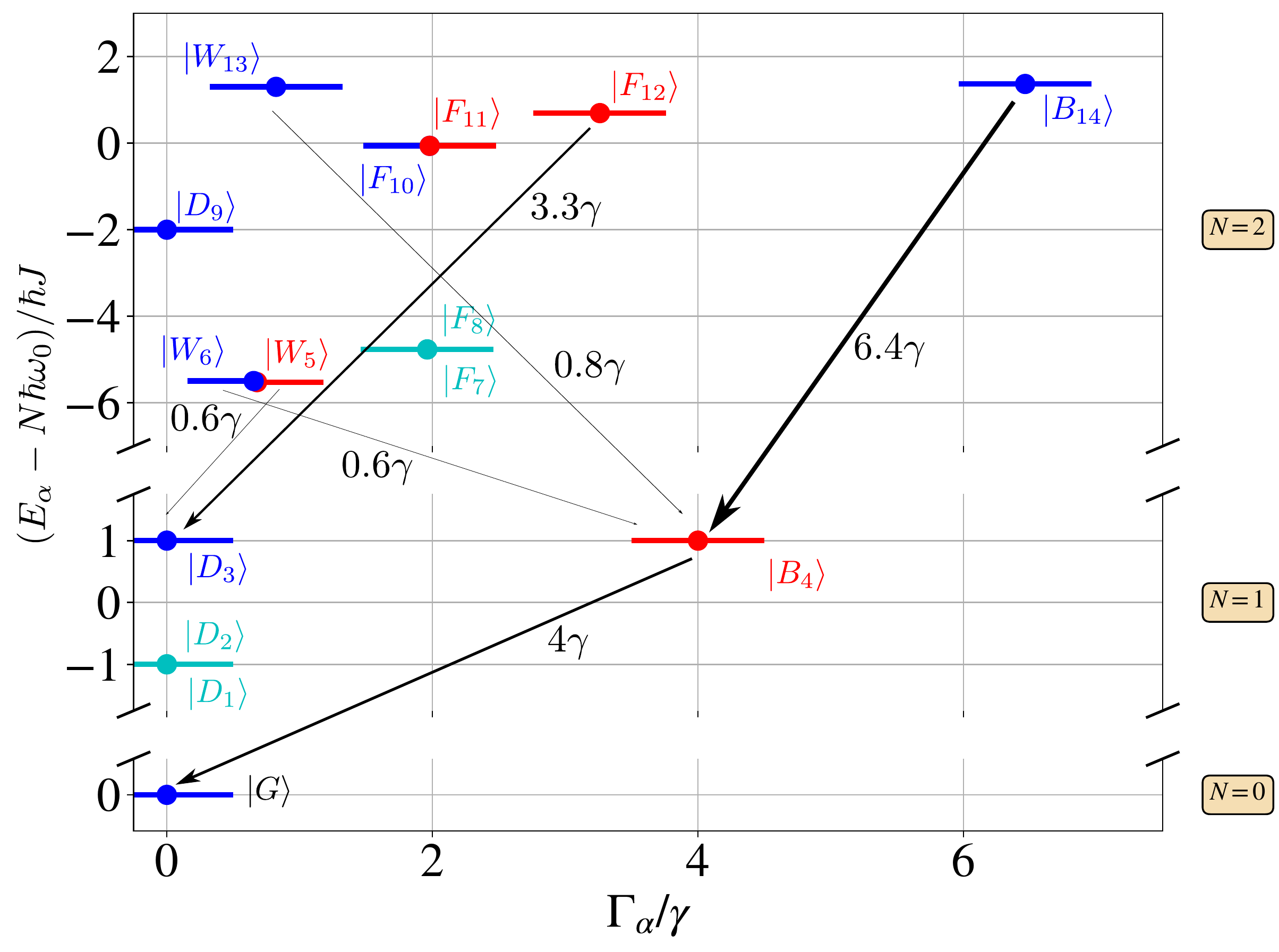}
    \caption{The complex valued eigenlevels $\lambda_\alpha=E_\alpha-i \hbar \Gamma_\alpha/2$ of the two-pair setup described by the Hamiltonian~\eqref{eq:two_pair_ham} in zero, one and two excitation manifolds marked with points. Red states are antisymmetric with respect to the exchange of the pair, and blue ones are symmetric. The grey ones do not possess pair-exchange symmetry. Pairs are separated by half of the wavelength corresponding to the transmon frequency~$\omega_0$. Black arrows display how the states connected to the global one-excitation states~$\ket{D_3}$ and~$\ket{B_4}$ decay through the waveguide, and their widths indicate the relative magnitude. The decay process by the collective decay operator $\cop_4$ of Eq.~\eqref{eq:b4} is antisymmetric, so only the decay events that change the symmetry are allowed. States~$\ket{W_{5,6}}$ and~$\ket{F_{7,8}}$ consist mainly of the states where two excitations occupy a single transmons, they do not exist in a qubit system. Additionally, the states~$\ket{F_{7,8,10,11}}$ decay to the local dark states~$\ket{D_{1,2}}$. States~$\ket{F_{7,8}}$ are degenerate, as well as~$\ket{F_{10,11}}$.}
    \label{fig:two_pairs_levels}
\end{figure}

As shown in Fig.~\ref{fig:levels_qubit_ho_transmon}, at low filling factors $N/L < 1/2$, the scaling of the collective decay rates is quite similar in transmon and qubit systems. Differences start to emerge at half filling, after which the decay rates of the brightest states and the Hilbert space dimensions in the qubit system start to decrease. In the bosonic systems they instead keep increasing. In small systems, containing two to four transmons, the collective bosonic effects should emerge already with two excitations. In this section we focus on a system consisting of two pairs of transmons, which is the simplest case of the array of pairs. Such a system is readily achievable also experimentally~\cite{Zanner2021}, which showcases the applicability and relevance of the presented theory. Here we use parameters that are close to the experimental ones except that the waveguide coupling~$\gamma$ has a larger value to highlight the collective bosonic effects, see Tab.~\ref{tab:exp_params}. 

The two-pair setup is also more versatile compared to the multiple pair setup, since the correlated decay between different pairs can be fully disabled here by tuning the transmon frequencies so that the pairs are an odd multiple of~$\lambda/4$ apart, which provides a coherent exchange interaction between the pairs instead. For multiple pairs this would remove the correlated decays between neighboring pairs, but to next-nearest pairs the correlated decay would again be at maximum, since the separation would be a multiple of~$\lambda/2$ instead.

The array of two transmon pairs is described by the effective Hamiltonian
\begin{align}\label{eq:two_pair_ham}
        \frac{\hop^{2+2}_{\rm eff}}{\hbar} &= \sum_{j=1}^4\left[\omega_j\nop_j-\frac{U}{2}\aop_j^\dag\aop_j^\dag\aop_j\aop_j\right]\\
        &+ J\left(\aop_1^\dag\aop_2 + \aop_3^\dag\aop_4 + \rm{h.c.}\right) -i \frac{\gamma}{2}\sum_{j,k=1}^4e^{i\omega_jt_{jk}}\aop_k^\dag\aop^{}_j,\notag 
\end{align}
where~$t_{jk}$ is the separation between sites~$j$ and~$k$. For sites forming the pair we have~$t_{jk} = 0$.
If all the transmons are in resonance,~$\omega_j=\omega_0$, and we assume that the pairs are now separated by half of the wavelength corresponding to the frequency, then based on Sec.~\ref{sec:coupled_system}, the one-excitation manifold is diagonalized by the operators
\begin{align}
    \dop_1 &= \frac{1}{\sqrt{2}}\Big(\aop_1-\aop_2\Big),
    & \Gamma_1&=0,\label{eq:b1}\\
    \dop_2 &= \frac{1}{\sqrt{2}}\Big(\aop_3-\aop_4\Big),
    & \Gamma_2&=0,\label{eq:b2}\\
    \cop_3 &= \frac{1}{2}\Big(\aop_1+\aop_2+\aop_3+\aop_4\Big),
    & \Gamma_3&=0,\label{eq:b3}\\
    \cop_4 &= -\frac{1}{2}\Big(\aop_1+\aop_2-\aop_3-\aop_4\Big),
    & \Gamma_4&=4\gamma,\label{eq:b4}
\end{align}
where~$\Gamma_\alpha$ are the corresponding decay rates. Here the lower index in the collective operators refers to the state, as opposed to Sec.~\ref{sec:coupled_system}. States~$\ket{D_1}=\dop_1^\dagger\ket{G}$ and~$\ket{D_2}=\dop_2^\dagger\ket{G}$ are the local dark states, and~$\ket{D_3}=\cop_3^\dag\ket{G}$ and~$\ket{B_4}=\cop_4^\dag\ket{G}$ are the collective dark and bright states. Numerically calculated eigenvalues of the Hamiltonian~\eqref{eq:two_pair_ham} are shown in Fig.~\ref{fig:two_pairs_levels} in zero, one and two excitation manifolds, with parameters given in Tab.~\ref{tab:exp_params}.

\begin{figure}[t]
    \centering
    \includegraphics[width=1.0\linewidth]{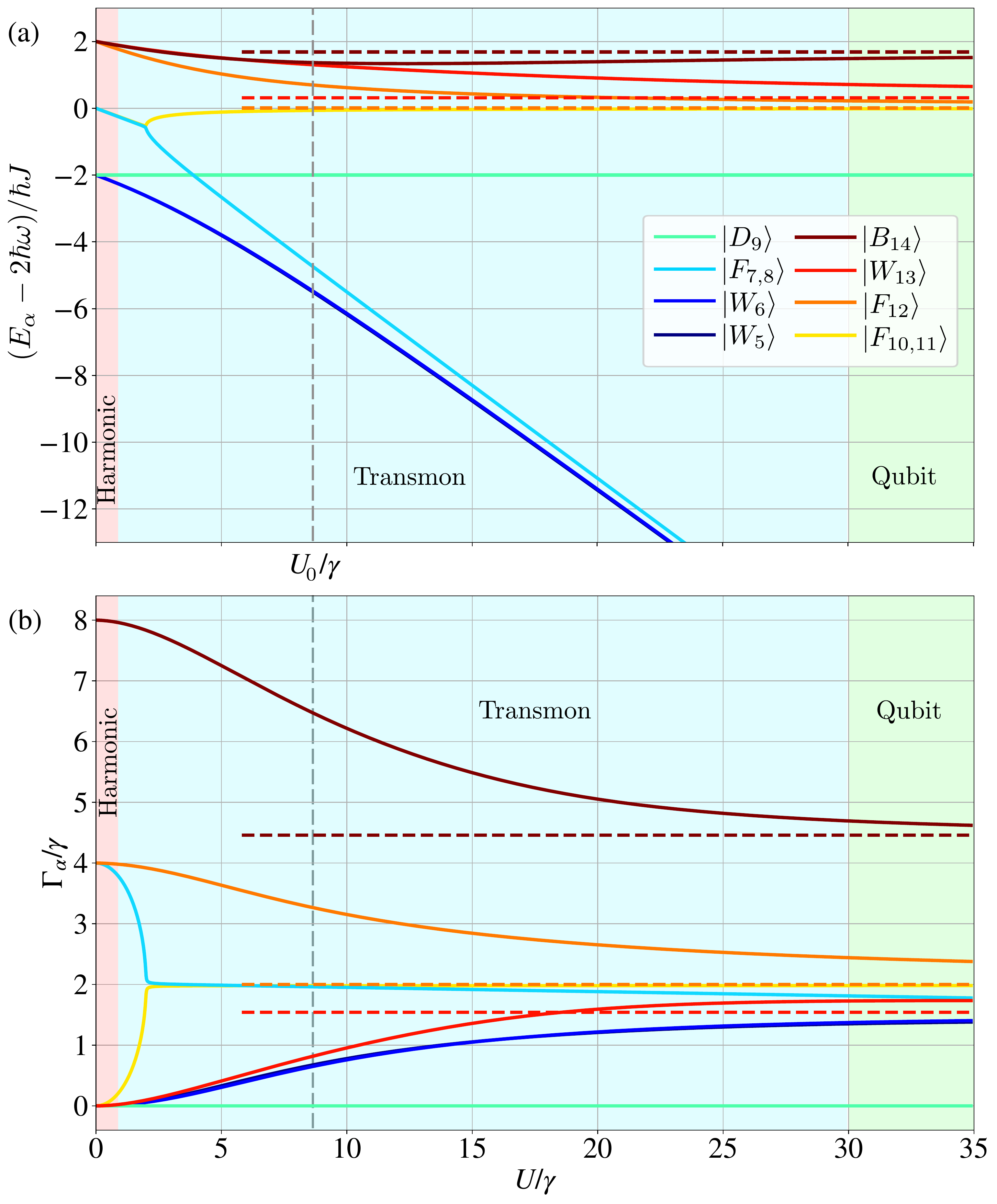}
    \caption{(a) Energies and (b) decay rates of the two-pair setup of Eq.~\eqref{eq:two_pair_ham} in the two-excitation manifold
    as a function of anharmonicity~$U$, other parameters are as in Tab.~\ref{tab:exp_params}. In the weak anharmonicity limit, transmons resemble harmonic  oscillators, but the collective complex eigenenergies rapidly deviate from that. However, the qubit eigenlevels (dashed horizontal lines) is achieved only at very large anharmonicities $U/\gamma \gtrsim 30$. Dashed vertical grey line describes the value of~$U/\gamma$ used in Fig.~\ref{fig:two_pairs_levels}, from which also the naming convention of the states is adapted.}
    \label{fig:qubitness}
\end{figure}

\begin{figure}[t]
    \centering
    \includegraphics[width=1.0\linewidth]{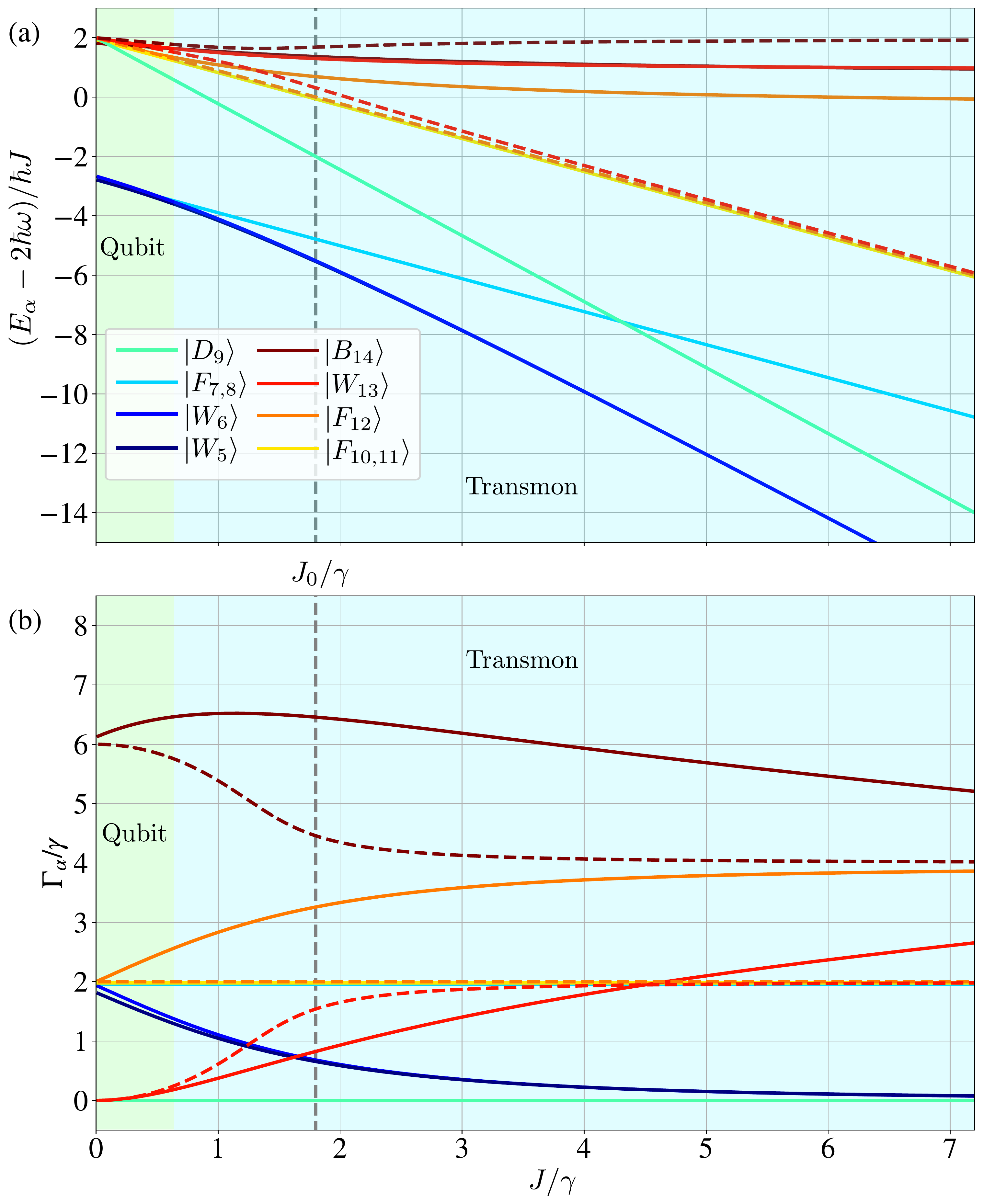}
    \caption{(a) Energies and (b) decay rates of the two-pair setup of Hamiltonian~\eqref{eq:two_pair_ham} in the two-excitation manifold
    as a function of the capacitive coupling~$J$. Other parameters are as in Tab.~\ref{tab:exp_params}. With weak coupling the system eigenvalues (solid curves) are close to the corresponding qubit system (dashed curves), but as~$J$ increases, they rapidly deviate. Dashed vertical grey line describes the value of~$J/\gamma$ used in Fig.~\ref{fig:two_pairs_levels}. Here we have kept the ratio~$U/\gamma = 8.72$ fixed and, because of such a large anharmonicity, there is no harmonic limit.}
    \label{fig:qubitness_fJ}
\end{figure}

As the number of excitations increases, the level structure develops a more complicated structure due to the interplay of the interaction and the collective decay. In the two-excitation manifold of Fig.~\ref{fig:two_pairs_levels}, there exists only one dark state~$\ket{D_9}=\ket{D_{1}}\otimes\ket{D_2}$, where both local dark states are excited. Lower in energy we find four states,~$\ket{W_{5,6}}$ and~$\ket{F_{7,8}}$, where W stands for weak and F for faint, referring to their moderate decay rates. These states are mostly made from Fock states where the two excitations occupy a single transmon, so they are affected by the anharmonicity more strongly than the other states, which decreases their energy. Despite their bosonic multi-excitation nature, the states $\ket{W_{5,6}}$ are almost dark. Moreover, because of the double occupancies, these states would not exist if the system was made from real qubits instead of transmons.  The remaining five states lie higher in energy. The states~$\ket{F_{10}}$ and~$\ket{F_{11}}$ are related to the local dark states, and the states~$\ket{W_{13}}$,~$\ket{F_{12}}$ and~$\ket{B_{14}}$ are mostly such that they contain two excitations in the collective dark state, one excitation in both collective states, and two excitations in the collective bright state, respectively, but due to the anharmonicity, they receive contributions also from the other states.

Different symmetries can be assigned to the collective eigenstates, but for us the most interesting one is the symmetry with respect to the exchange of the pairs, defined by the operator~$\hat P = \ket{n_3n_4n_1n_2}\bra{n_1n_2n_3n_4}$, where~$n_j$ is the number of excitations at the~$j$th site. The pair-exchange symmetry defines which decay processes are possible and which kind of collective drive is needed to couple  states and to induce transitions between them.~If the pair-exchange operator leaves a state intact,~$\hat P\ket{\alpha}=1\ket{\alpha}$, the state is symmetric, and if the state becomes itself with a sign change,~$\hat P\ket{\alpha}=-1\ket{\alpha}$, it is antisymmetric. Not every state has this symmetry, for example the local states~$\ket{D_1}$ and~$\ket{D_2}$ in the one-excitation manifold, since they contain excitation in one pair only. On the other hand, Eqs.~\eqref{eq:b3} and~\eqref{eq:b4} show that the global states~$\ket{D_3}$ and~$\ket{B_4}$ are symmetric and antisymmetric, respectively. The decay process through the waveguide is antisymmetric through the decay operator $\cop_4=-(\aop_1+\aop_2-\aop_3-\aop_4)/2$, which means that it connects states with the opposite pair-exchange symmetries. This is visible in Fig.~\ref{fig:two_pairs_levels}, where the symmetric states~$\ket{W_{13}}$ and~$\ket{B_{14}}$ decay to the antisymmetric state~$\ket{B_4}$, which further decays to the symmetric ground state~$\ket{G}$. Similarly, the antisymmetric state~$\ket{F_{12}}$ decays to the symmetric state~$\ket{D_3}$, which cannot decay further, because the ground state has the same symmetry.

In Fig.~\ref{fig:qubitness} we plot the eigenvalues of the two excitation manifold of the Hamiltonian~\eqref{eq:two_pair_ham} as a function of anharmonicity~$U$ in order to observe the transition from a harmonic to a qubit system. Fig.~\ref{fig:qubitness}(a) shows the energies and Fig.~\ref{fig:qubitness}(b) the corresponding decay rates. Corresponding values for qubits are shown as dashed horizontal lines, and dashed vertical line denotes the parameters at which the results in Fig.~\ref{fig:two_pairs_levels} are calculated. The states are labeled according to these values, although their radiative properties change as a function of anharmonicity, as is evident from Fig.~\ref{fig:qubitness}(b). The system rapidly deviates from the harmonic description (red region) as the anharmonicity increases. Especially we note that the four states~$\ket{W_{5,6}}$ and~$\ket{F_{7,8}}$ containing the double excited Fock states rapidly decrease in energy as a function of anharmonicity, and so they become detuned from the qubit space. The states~$\ket{F_{10,11}}$, which lie higher in energy, form an exceptional point~\cite{Eleuch2014} with the states~$\ket{F_{7,8}}$ at~$U=2\gamma$. For smaller~$U$ these states are degenerate in energy, and they form dark and bright states. For larger~$U$, their decay rates become degenerate, but their energies deviate.

Finally, the three states~$\ket{F_{12}}$,~$\ket{W_{13}}$ and~$\ket{B_{14}}$, which have the highest energies, begin as bosonic collective states with the decay rates~0,~$4\gamma$ and~$8\gamma$ and degenerate energies. As the anharmonicity increases, they deviate in energy and slowly converge towards the qubit energies and decay rates. The qubit limit sketched in Fig.~\ref{fig:qubitness} corresponds to~$U/2\pi=\SI{750}{\mega\hertz}$, which is much larger than the typical value for transmon anharmonicity. Thus, in practice, with the parameters used in Fig.~\ref{fig:qubitness}, the transmon system cannot be approximated as a qubit system.

There are actually three parameters~$U$,~$\gamma$ and~$J$, whose interplay affects the behavior of the system. In Fig.~\ref{fig:qubitness} the ratio~$J/\gamma$ is kept fixed. Altering the value of~$J$ also affects the system, as shown in Fig.~\ref{fig:qubitness_fJ}. The transmon system is more qubit-like for smaller~$J$. However, especially the decay rates of the high-energy states~$\ket{F_{12}}$,~$\ket{W_{13}}$ and~$\ket{B_{14}}$ require very low~$J$ in order to be close to the qubit values. On the other hand, the states~$\ket{F_{10,11}}$ are very close to qubit ones at the shown range. In conclusion, there exist a wide range of experimentally realizable parameters with which the transmon system resembles neither a harmonic nor a qubit system. Especially, with multiple transmons and excitations, the widely used two-level approximation is actually in many cases not applicable, but the anharmonic model should be used instead.

\subsection{Detuning between the pairs}
\begin{figure}[t]
    \centering
    \includegraphics[width=1.0\linewidth]{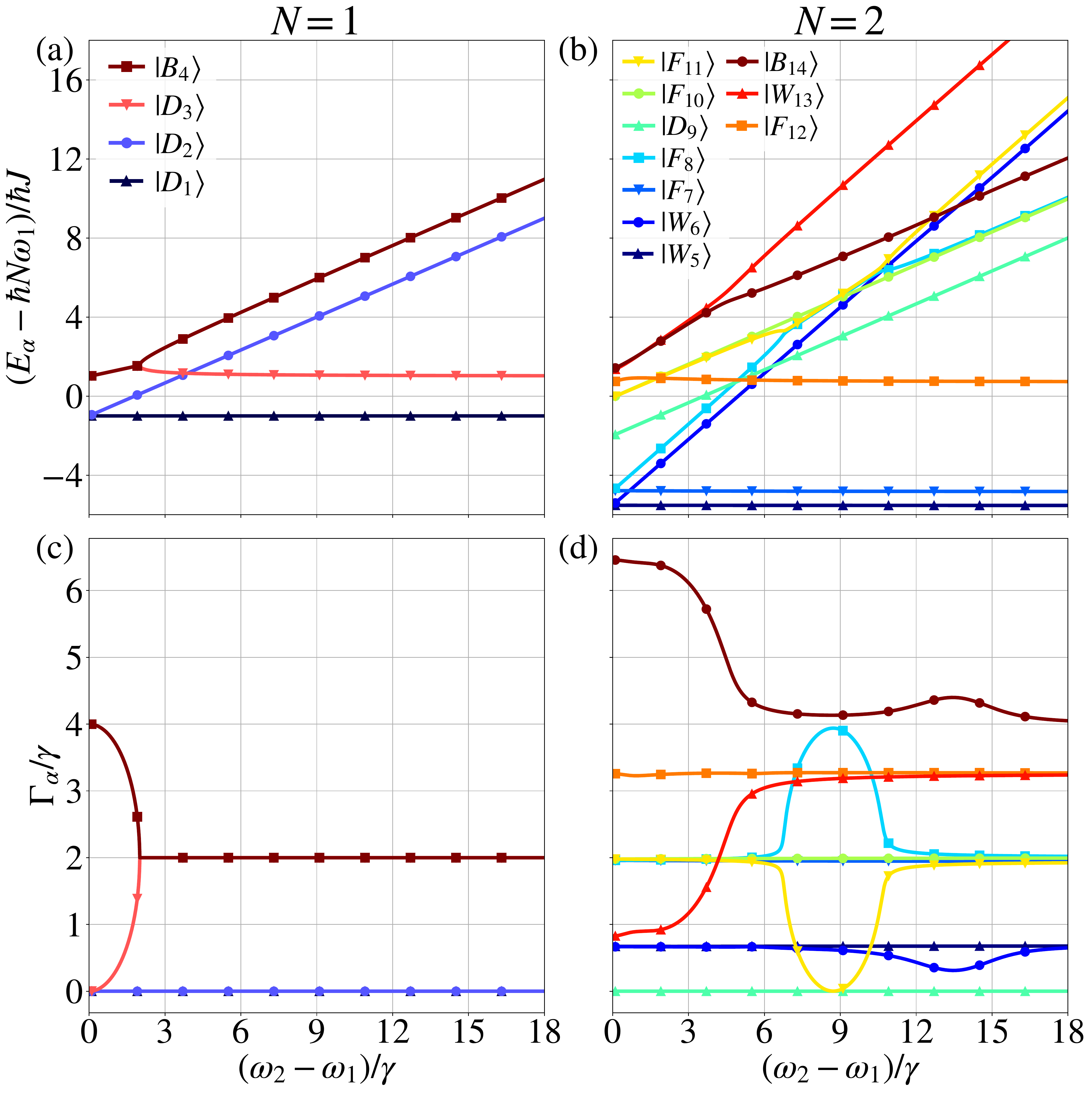}
    \caption{Energies (top row) and the corresponding decay rates (bottom row) as a function of detuning between the transmon pairs of the states in one (left column) and two (right column) excitation manifolds. The pair separation is such that they are always half of the wavelength of their average frequency apart. Points at which the decay rates separate and energies become degenerate are called exceptional points. System parameters are given in Tab.~\ref{tab:exp_params}, except for~$\omega_2$.}
    \label{fig:states}
\end{figure}
In the scenario above we assumed that the pairs are in resonance. This caused the collective decay and the emergence of bright and dark states. Assuming that the first pair has frequency~$\omega_1$ and the other pair~$\omega_2$, the complex eigenvalues of the collective global states of the effective Hamiltonian in the one-excitation manifold are
\begin{equation}
    \lambda_{3,4}/\hbar = \frac{\omega_1+\omega_2}{2}+J-i\gamma\pm\frac{1}{2}\sqrt{\left(\omega_1-\omega_2\right)^2-
    4\gamma^2}.
\end{equation}
The states~$\ket{D_1}$ and~$\ket{D_2}$ are local, and thus their behavior is not affected by the detuning, unlike the two collective states. If the detuning between the transmons is larger than~$2\gamma$, the two collective states have the same decay rate, but their energy is different. At detuning~$2\gamma$ the eigenvalues become degenerate in energy and decay rate, since the square root vanishes. For a detuning less than~$2\gamma$, the argument in the square root is negative, so the term gives an imaginary part to the eigenvalues, which modifies their decay rates. The states are degenerate in energy, but their decay rates start to deviate. At the resonance the other state is completely dark, while the other obtains a maximal decay rate~$4\gamma$. Thus, the system has an exceptional point~\cite{Eleuch2014,Ashida2020} at~$|\omega_1-\omega_2|=2\gamma$ between the states~$\ket{D_3}$ and~$\ket{B_4}$. This behavior is shown in Fig.~\ref{fig:states}(a) for energies and in Fig.~\ref{fig:states}(c) for decay rates as a function of pair detuning for the states in the one-excitation manifold (only the positive~$x$ axis is shown). 

The behavior of the two-excitation states is shown in Figs.~\ref{fig:states}(b) and (d) for the energy and decay rates, respectively. With the chosen parameters the two-excitation manifold contains multiple regions where certain states exhibit exceptional point -like behavior, most notably between states~$\ket{F_{8}}$ and~$\ket{F_{11}}$. These occur anharmonicity away from the resonance, so they are characteristic for anharmonic oscillators.  However, unlike in the one-excitation manifold, here the decay rates and energies do not become strictly degenerate, due to the effects of anharmonicity and capacitive coupling, in addition to the frequency detuning. Similar behavior occurs also between states~$\ket{B_{14}}$ and~$\ket{W_{13}}$, and more weakly between the states~$\ket{B_{14}}$ and~$\ket{W_{6}}$. However, in these cases the decay rates do not coalesce due to the effect of capacitive coupling and anharmonicity~\cite{Eleuch2014}.

\section{Observables of the collective spectrum}\label{sec:spectroscopy}
In this section we discuss four possible experimentally realizable observables that could be used for studying the collective phenomena of transmon arrays in a waveguide. We introduce superradiant radiation bursts, transmission spectra, emission spectra and direct spectroscopy of the second excitation manifold. Especially we focus on the features that distinguish the bosonic collective phenomena from those of qubit arrays.

\subsection{Superradiant burst}
\begin{figure}
    \centering
    \includegraphics[width=1.0\linewidth]
    {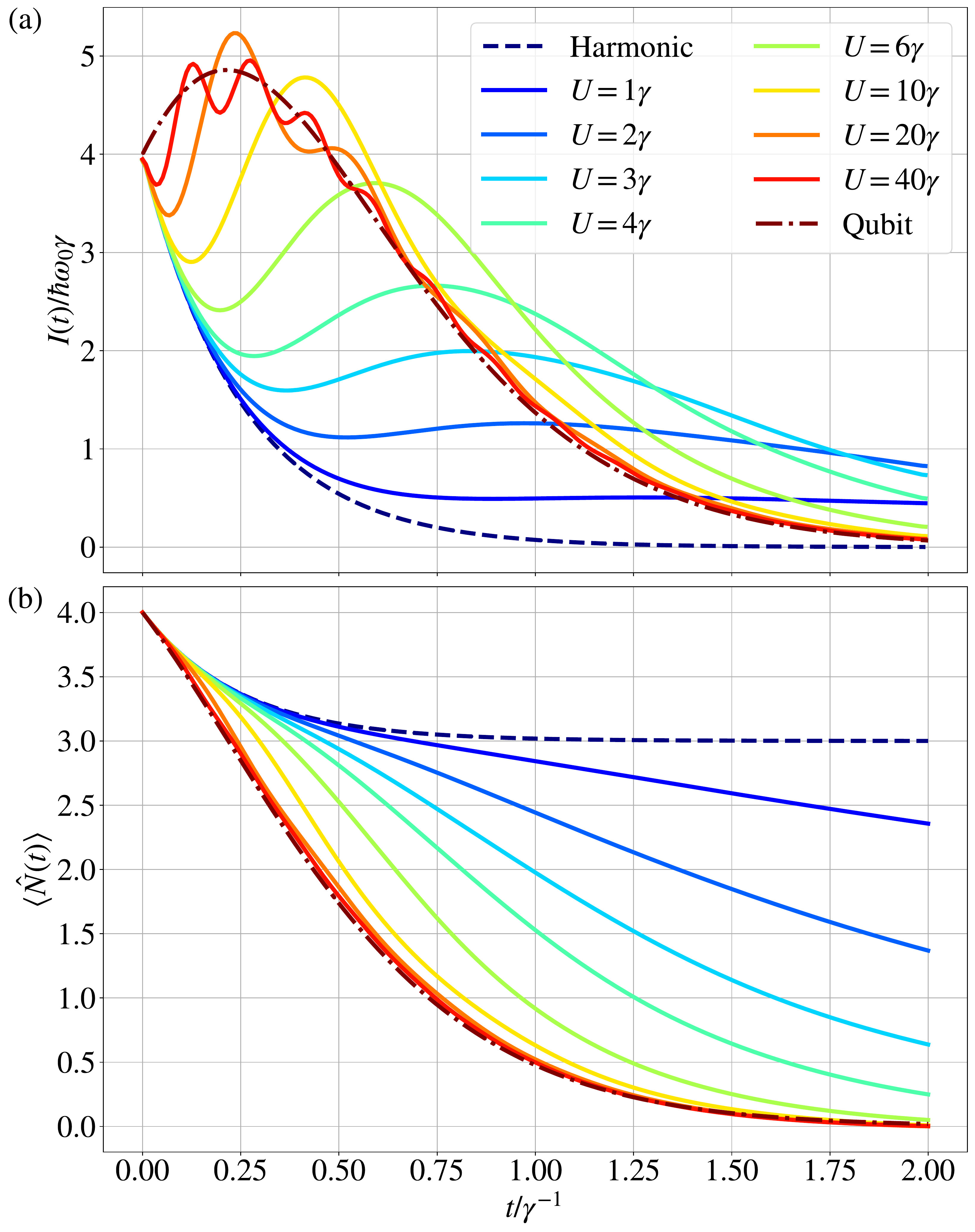}
    \caption{(a) Intensity of the outcoming radiation as a function of time, defined as $I(t) =-\hbar\omega_0 d\braket{\ntot(t)}/dt$, for different anharmonicities. (b) The total occupation $\braket{\ntot(t)}$ as a function of time. The system contains four uncoupled sites at a wavelength distance apart, and the initial state is $\ket{1111}$. For harmonic oscillators the decay of the population is exponential, which results also in an exponential decay of the intensity. In the other limit the sites are qubits. In this case the intensity initially increases, obtains a maximum, and starts to decrease. This is the superradiant burst of emission. The transmon behavior is in between these two cases. The system initially behaves as harmonic with exponentially decreasing intensity, but after a while the intensity increases temporarily. As the anharmonicity increases, the burst occurs earlier and with larger intensity, approaching the qubit solution. The behavior in bosonic systems depends on the initial state. If the system is initially in the brightest state, also transmons would decay exponentially, and no superradiant burst would occur. Note that for bosonic systems the initial state is not an eigenstate of the effective Hamiltonian, but instead some linear combination of them.}
    \label{fig:dickeburst}
\end{figure}

The superradiant burst is a fundamental characteristic of Dicke superradiance of qubits~\cite{masson2021universality, Slama2007, Scheibner2007}. The burst is observed when a collectively decaying array of $L$ qubits is prepared in the highest excited state, that is, $\ket{11\ldots11}$. Referring to Fig.~\ref{fig:levels_qubit_ho_transmon}(a), this state decays by rate $\gamma L $ to a state that further decays with a larger rate. At half-filling the decay rates start to decrease and the final state is the ground state, see the green dashed line in Fig.~\ref{fig:levels_qubit_ho_transmon}(a). This decay path generates a burst of radiation, see Fig.~\ref{fig:dickeburst}(a) where we show the intensity of radiation, defined as a time derivative of the total occupation of the system,~$I(t) = -\hbar\omega_0d\braket{\ntot(t)}/dt$~\cite{Breuer}.~The situation changes drastically by considering the same scenario with an array of collectively decaying harmonic oscillators. The harmonic oscillator system does not show signs of superradiant burst, see Fig.~\ref{fig:dickeburst}(a).~First of all, it is not possible to define uniquely the highest excited state due to the bosonic excitation statistics, thus, the initial state $\ket{11\ldots11}$ is a superposition of the collective states $\ket{m_1,m_2, \ldots, m_L}$ with the total excitation number $\braket{\ntot}=L$ and only the states with $m_L\neq 0 $ are decaying. Furthermore, from the superposition only the states with $\ket{m_1, m_2, \ldots, m_L=k}$ decay exponentially to a final state that is a dark state $\ket{m_1,m_2, \ldots,m_L=0}$ whose total occupation number is $\braket{\hat N}=L-k$, see the diagonal black arrays in  Fig.~\ref{fig:levels_qubit_ho_transmon}(c). 

A transmon array shows behavior that is in between the pure qubits and harmonic oscillators. For weak anharmonicity $U/\gamma\lesssim 5$, the transmon arrays is closer to that of harmonic oscillators, but as the anharmonicity increases, a peak in the intensity starts to emerge. For large anharmonicity~$U/\gamma\gtrsim 10$, the intensity approaches the qubit solution with additional oscillations~\cite{Delanty2011}. Initially, a transmon array decays fast resembling an array of harmonic oscillator and later shows a burst of radiation that is delayed compared to pure qubit case. Qualitatively we can understand this so that the initial state $\ket{11\dots11}$ is a superposition of the collective eigenstates of the transmon array. The collective states that most resemble those of an array of harmonic oscillator have the largest decay rate and thus decay the fastest. The remaining states, which are rendered similar to those of a qubit array by the interaction term, show a characteristic superradiant burst. This behavior repeats as the system loses excitations, which results in the oscillatory behavior visible with large anharmonicity~$U$ in Fig.~\ref{fig:dickeburst}. Notice that for bosonic systems the exact form of the radiation depends on the initial state.

\subsection{Probing through the waveguide}
\begin{figure*}[ht]
    \centering
    \includegraphics[width=1.0\linewidth]
    {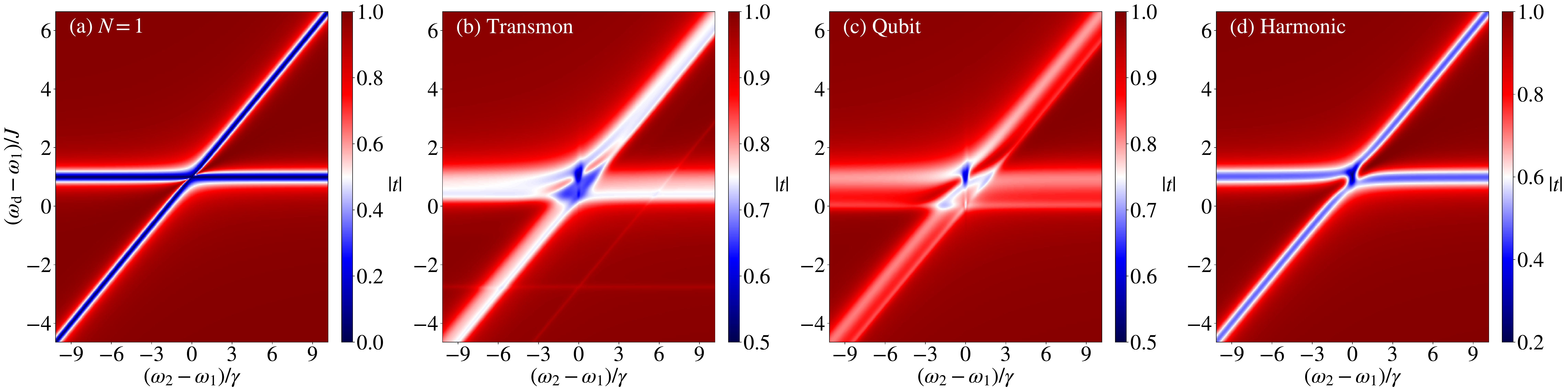}
    \caption{Waveguide transmission amplitude in a two-pair system as a function of detuning $\omega_2-\omega_1$ between the pairs, and the probe frequency $\omega_d$. 
    In (a) the driving power is~$P/2\pi=\SI{0.7}{\kilo\hertz}$, which excites only the one excitation eigenstates, and thus the results are the same for all three models. In the remaining ones the driving power is~$P/2\pi=\SI{22}{\mega\hertz}$. Other system parameters are given in Tab.~\ref{tab:exp_params}. The system consists of (b) transmons, (c) qubits, and (d) harmonic oscillators.}
    \label{fig:transmission_twopairs}
\end{figure*}
The superradiant burst on itself does not give information about the individual eigenstates of the system. These can instead be studied by using suitable drives to excite them, and then observing their decay. Let us first consider a situation where the system of two transmon pairs, discussed in Sec.~\ref{sec:twopairs}, is driven through the waveguide. The Hamiltonian is time dependent, but since there is only one frequency involved, one can switch to a frame rotating with the driving frequency and remove the time-dependence by doing the rotating wave approximation. This gives the Hamiltonian
\begin{align}
        \frac{\hop_{\rm probed}}{\hbar} &= 
        \frac{\hop_{2+2}}{\hbar} - \omega_{\rm d}\ntot
        +\sum_{mj,nk}J_{mj,nk}\spop^{nk}\smop^{mj} \notag \\
        &+\sum_{mj}\left(\widetilde d_{mj}\smop^{mj} + 
        {\widetilde d}^*_{mj}\spop^{mj}\right), \label{eq:probe_ham}
\end{align}
where~$\omega_{\rm d}$ is the frequency of the drive, and
\begin{equation}
    \widetilde d_{mj} =i\sqrt{\frac{P\gamma_{mj,mj}}{2\hbar\omega_{mj}}}e^{i\omega_{\rm d}z_j/c} \label{eq:dtilde}
\end{equation}
is the amplitude of the coherent driving where $z_j$ is the coordinate of the site $j$, see App.~\ref{sec:above_and_below} for details.
The master equation describing the system dynamics is then
\begin{align}
        \frac{d\dens}{dt} &= -\frac{i}{\hbar}\Big[\hop_{\rm probed},\dens\Big]
        +\sum_{j}\kappa\left(\aop_j\dens\aop_j^\dag-\frac{1}{2}\big\{\aop_j^\dag\aop_j,\dens\big\}
        \right) \nonumber\\
        &+\sum_{mj,nk}\gamma_{mj,nk}\left(\smop^{mj}\dens\spop^{nk} - \frac{1}{2}
        \big\{\spop^{nk}\smop^{mj},\dens\big\}\right),\label{eq:probe_me}
\end{align}
where~$\kappa$ is the intrinsic dissipation rate of the transmons, which we here denote as the bulk dissipation rate to distinguish it from the dissipation $\gamma$ via the waveguide, $\kappa\ll\gamma$. At weak power, the driving does not affect the energy levels of the transmon system, but only induces transitions between them. The interplay of the driving and dissipation eventually leads to a steady state. Without bulk dissipation, the steady state can depend on the initial state, which leads to a multiple possible steady states. This happens because the system can have multiple dark states, so any arbitrary initial population in those also remains there. On the other hand, bulk dissipation gives additional decay rates to all states, and thus also dark states decay, and there exists only one steady state. In the numerical simulation we solve the steady state of the master equation in Eq.~\eqref{eq:probe_me} and calculate the transmission of radiation~$|t|^2$ in it, as discussed in App.~\ref{sec:master_equation_io}. If all the radiation comes through, there was no state that could have been excited by the drive. If some fraction of the radiation is lost, it was absorbed by the system, resulting in an excitation of a state.

In the limit of weak driving, the transmission can be solved analytically. We denote $\Delta=\omega_1-\omega_2$ as the detuning between the pairs and the driving frequency is detuned by $\delta=\bar\omega-\omega_{\rm d}$ from the average pair frequency $\bar\omega=(\omega_1+\omega_2)/2$. When the pairs are separated by a distance~$\lambda/2$, we find the transmission
\begin{equation}\label{eq:trans_12}
    |t|^2 = \frac{\left[(\delta+J)^2-\frac{\Delta^2}{4}\right]^2}
	{\left[(\delta+J)^2-\frac{\Delta^2}{4}\right]^2+4\gamma^2(\delta+J)^2},
\end{equation}
where we have neglected the bulk dissipation~$\kappa$. Transmission vanishes at~$\Delta=\pm2(\delta+J)$, i.e., when the probe frequency is~$\omega_{\rm d} = \omega_{1,2}+J$, which are the transition frequencies of the bare qubit system, in the absence of the waveguide interactions. This means that the transmission probes the eigenstates of the Hermitian Hamiltonian, not those of the effective non-Hermitian one. Because of this we do not see the emergence of the exceptional points in the transmission spectrum, see Fig.~\ref{fig:states}(a). For example, at the exceptional points at detuning~$\Delta\pm 2\gamma$ the collective states have degenerate energy $\bar\omega+J$. Probing at this frequency gives~$\delta=-J$, which results in perfect transmission~$|t|^2=1$, except at~$\Delta=0$ at which the transmission vanishes. However, at~$\Delta=0$ the width of the Lorentzian at half maximum, centered around~$\delta+J$, is~$4\gamma$, which is the bright state decay rate. The features described by Eq.~\eqref{eq:trans_12} are accurately captured in the full numerical simulations shown in Fig.~\ref{fig:transmission_twopairs}(a). Here the driving amplitude is weak, so that only the states in the one-excitation manifold are excited. The one-excitation manifold contains four states, but we see only two spectral lines. Both local pairs have two states at energies~$\omega_i\pm J$. The corresponding states are the same as local bright and dark states. Because of this, only the states at energies~$\omega_{1,2}+J$ are visible in the transmission, since they are the bright states and thus couple to the waveguide field.

All three models, transmon, qubit and harmonic oscillator, are identical in the one-excitation manifold. The differences emerge in the two-excitation manifold, which can be studied, e.g., by increasing the power of the probe. In Fig.~\ref{fig:transmission_twopairs}(b) we show the transmission with larger driving for a system of transmons. We now observe four additional states corresponding to two photon transitions between the ground state and the two-excitation manifold, two of which are low in frequency due to the anharmonicity arising from multiple occupations in individual sites. For comparison, the results for qubit and harmonic oscillator systems are shown in Figs.~\ref{fig:transmission_twopairs}(c) and (d), respectively. In the harmonic oscillator system all the transitions occur at the same frequency, so no additional spectral features become visible. In qubit system, on the other hand, we observe similar features as with transmons, but the bosonic states in low frequency do not exist.

\subsection{Spectral density}
\begin{figure*}[ht]
    \centering
    \includegraphics[width=1.0\linewidth]
    {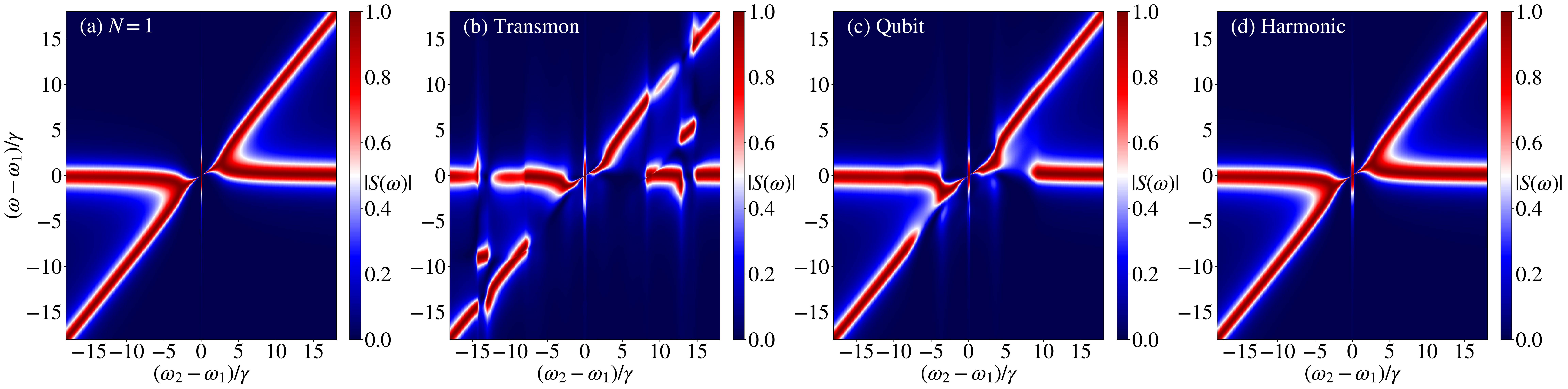}
    \caption{Magnitude of the spectral density~$|S(\omega)|$ of the out-coming radiation in a two-pair system as a function of the detuning between the pairs~$\omega_1-\omega_2$, and the frequency of the out-coming radiation $\omega$. In (a) the driving power is  ~$P/2\pi=\SI{0.7}{\kilo\hertz}$, which excites only the one excitation eigenstates. In the remaining ones the driving power is~$P/2\pi=\SI{22}{\mega\hertz}$. Other system parameters are given in Tab.~\ref{tab:exp_params}. The system consists of (b) transmons, (c) qubits, and (d) harmonic oscillators. We have separately normalized each~$|S(\omega)|$ at different detunings~$\Delta=\omega_1-\omega_2$ for better visibility. The main feature is the emergence of exceptional points around the resonance. The system of transmons exhibits the most complicated spectrum of the three models, with additional features occurring at frequencies at which the system has exceptional point -like behavior in the two excitation manifold, as shown in Fig.~\ref{fig:states}.}
    \label{fig:spectral_density_twopairs}
\end{figure*}
The transmission measurement described in the previous section probes the local eigenstates. The collective eigenstates of the effective non-Hermitian Hamiltonian can be studied by using the power spectrum of the output field, defined as
\begin{equation}\label{eq:power_spectrum}
    S_{\rm L/R}(\omega) = \int_{-\infty}^\infty dt e^{i\omega t}
    \braket{\aop_{\rm L/R}^{{\dag\rm out}}(t)\aop_{\rm L/R}^{\rm out}(0)},
\end{equation}
where L and R refer to left and right moving excitations, and~$\omega$ is the frequency at which the system radiates,~$\omega=0$ corresponding to the probe frequency. Here we calculate the power spectrum by driving the system coherently with the amplitude $\braket{\aop^{\rm in}}$, see Eq.~\eqref{eq:ain}, until it reaches a steady state, then turn off the drive and let the system decay and radiate. The outcoming radiation at different frequencies is then given by Eq.~\eqref{eq:power_spectrum} where the left moving mode~$\aop_{\rm L}^{\rm out}$ is defined in Eq.~\eqref{eq:aout}. We sweep over the pair detuning~$\Delta=\omega_1-\omega_2$, and drive the system coherently with the average frequency of the two pairs,~$\omega_{\rm d} = (\omega_1+\omega_2)/2$.  For a weak probing power and~$\lambda/2$ separation of the pairs, we obtain the analytical formula
\begin{equation}\label{eq:spectral_density}
    |S_{\rm L,R}|^2 = \frac{4\gamma^2\braket{\aop^{\rm in}}^4}
    {\left((\omega-J-\frac{\Delta}{2})^2-\frac{\Delta^2}{4}\right)^2+4(\omega-J-\frac{\Delta}{2})^2\gamma^2},
\end{equation}
where we have assumed that the system is driven from the left only, and again ignored the bulk dissipation~$\kappa$. 

Result in Eq.~\eqref{eq:spectral_density} agrees well with the numerical simulations shown in Fig.~\ref{fig:spectral_density_twopairs}(a), in which we observe two energy levels that coalesce into one at the exceptional point. Curiously, the spectral density shows the exceptional points already at~$\Delta=\pm2\sqrt{2}\gamma$, i.e., at a slightly larger detuning. This is in agreement with Eq.~\eqref{eq:spectral_density}. The linewidth of the bright state is visible only on resonance, where we observe a Lorentzian with width~$4\gamma$. For weak probe power, the results are the same for all three models. As the power is increased, also the two-photon manifold becomes excited and starts to radiate. For harmonic oscillators in Fig.~\ref{fig:spectral_density_twopairs}(d) this only affects the features near resonance: the large linewidth of the bright state is now visible also slightly off-resonance. 

In qubit and transmon systems we instead observe cuts in the spectral lines, which tells that one of the states radiates more strongly than the other one. Most importantly, the additional exceptional points in the transmon system between the states~$\ket{F_{7(8)}}$ and~$\ket{F_{10(11)}}$, centered around~$\Delta=\pm U\approx8.7\gamma$ (see section~\ref{sec:twopairs}), show up weakly. We also see some features around~$\Delta\approx\pm 13\gamma$, which can be attributed to the enhancement and suppression of the decay rates of states~$\ket{B_{14}}$ and~$\ket{W_6}$, see Fig.~\ref{fig:states}.

\subsection{Pulsed excitation of the two-excitation manifold}
When driving the array through the waveguide, the collective drive has a symmetry set by the separation of the sites, according to Eq.~\eqref{eq:dtilde}, which means that it has the same symmetry as the global bright state~$\ket{B_4}$. To go beyond, in Ref.~\cite{Zanner2021}, we experimentally demonstrated on-site driving through waveguide sideports with tunable frequency, as well as local amplitudes and phases. Such a drive can be modeled with the Hamiltonian
\begin{equation}
        \frac{\hop_{\rm d}(t)}{\hbar} = 2\cos\left(\omega_{\rm p} t\right)\sum_j
        A_j\left(e^{i\phi_j}\aop_j + e^{-i\phi_j}\aop_j^\dag
        \right),
\end{equation}
where~$\omega_{\rm p}$ is the on-site driving frequency,~$A_j$ are the local amplitudes, and $\phi_j$ are the local phases. Assuming that the phases within pairs are the same, but there is a phase difference~$\phi$ between the pairs, and further that the amplitudes are the same for all sites, we can write the driving Hamiltonian in terms of the global collective operators, defined in Eqs.~\eqref{eq:b1}-\eqref{eq:b4}. Performing also the rotating wave approximation results in
\begin{equation}
    \frac{\hop_{\rm d}}{\hbar} = A\left[\left(1+e^{i\phi}\right)\cop_3 + \left(1-e^{i\phi}\right)\cop_4 + \rm{h.c.}\right].
\end{equation}
With such a drive only the global states~$\ket{D_3}$ and~$\ket{B_4}$ can be excited from the ground state. The phase difference~$\phi$ determines the symmetry of the drive. Symmetric and antisymmetric drives always couples states with the same and opposite symmetries, respectively, see Fig.~\ref{fig:two_pairs_levels}. Clearly for even multiples of~$\pi$, the drive is symmetric, and for odd multiples it is antisymmetric.

The symmetries of the global states in the one-excitation manifold of the two-pair setup provide a scheme for probing the two-excitation manifold. First, one can employ a suitable Rabi pulse, which excites the long lived dark state~$\ket{D_3}$ from the ground state. Then one can apply another pulse with a different frequency and symmetry, which can excite one of the two-excitation states. Some of these states decay to the one-excitation bright state~$\ket{B_4}$, which further rapidly decays to the ground state, see the decay channels illustrated in Fig.~\ref{fig:two_pairs_levels}. 

\begin{figure*}[ht]
    \centering
    \includegraphics[width=1.0\linewidth]{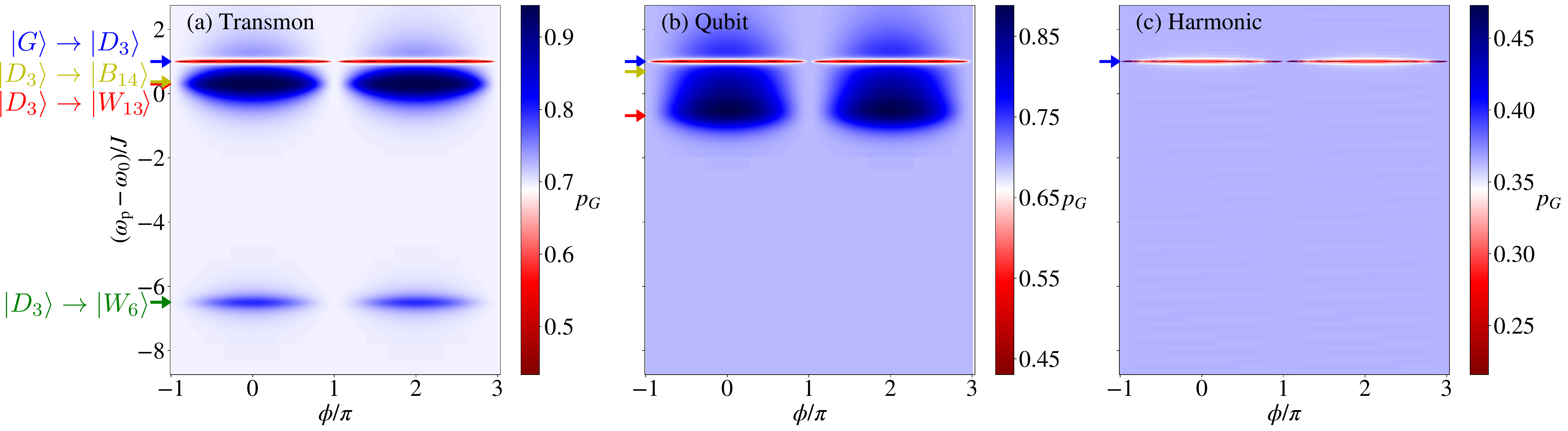}
    \caption{Ground state population after a symmetric Rabi pulse and a subsequent spectroscopy pulse with altering phase difference between the pairs~$\phi$ and frequency~$\omega_{\rm p}$ for a two pair setup of (a) transmons, (b) qubits, and (c) harmonic oscillators. The transitions are denoted by arrows and they refer to the eigenlevels of Fig.~\ref{fig:two_pairs_levels}. The parameters are  $T_{\rm Rabi} = \SI{240}{\nano\second}$, $\mu_{\rm Rabi} =T_{\rm Rabi}/2$ and $A_{\rm Rabi}/2\pi = \SI{4}{\mega\hertz}$ for the Rabi pulse and $T_{\rm spec} = \SI{1200}{\nano\second}$, $\mu_{\rm spec} = T_{\rm Rabi} + T_{\rm spec}/2$ and $A_{\rm spec}/2\pi = \SI{1}{\mega\hertz}$ for the spectroscopy pulse. For both pulses~$\sigma = T/6$. Other system parameters are given in Tab.~\ref{tab:exp_params}.}
    \label{fig:transmission_pulse_probe}
\end{figure*}

The system is driven with two consecutive drive fields, in a pulsed fashion. Therefore one can reduce the time-dependence by switching to a frame rotating with the frequency of the drive and solve the dynamics numerically. Once another drive is applied, one has to change the Hamiltonian and switch to another frame. The amplitudes of the pulses are time dependent,
\begin{equation}
    A(t) = Ae^{-(t-\mu)^2/(2\sigma^2)},
\end{equation}
where~$A$ is the amplitude,~$\mu$ is the time instance at which the pulse is at maximum, and~$\sigma$ is the width of the pulse. This means that the rotating wave approximation does not remove the time dependence completely. However, it makes solving the system numerically more stable. In the simulation we then have two Hamiltonians,
\begin{equation}
    \begin{aligned}
    \frac{\hop_{1}(t)}{\hbar} &= \frac{\hop_{\rm 2+2}}{\hbar}- \omega_{\rm Rabi}\ntot
    +\sum_{mj,nk}J_{mj,nk}\spop^{nk}\smop^{mj}\\
    &+A_1(t)\Big(\aop_1+\aop_2+\aop_3+\aop_4+\rm{h.c.}\Big),\label{eq:H1}
    \end{aligned}
\end{equation}
\begin{equation}
    \begin{aligned}
    \frac{\hop_{2}(t)}{\hbar} &= \frac{\hop_{\rm 2+2}}{\hbar} - \omega_{\rm p}\ntot
    +\sum_{mj,nk}J_{mj,nk}\spop^{nk}\smop^{mj}\\
    &+A_2(t)\Big[e^{i\phi}\big(\aop_1+\aop_2\big)+\aop_3+\aop_4+\rm{h.c.}\Big],
    \label{eq:H2}
    \end{aligned}
\end{equation}
where we have performed the rotating wave approximation in each. The time evolution is governed by the master equation of Eq.~\eqref{eq:probe_me} with~$\hop_{\rm probed}$ replaced by Eqs.~\eqref{eq:H1} and~\eqref{eq:H2}. In the numerical simulation we initially set the system to the ground state~$\ket{G}$ and calculate the time evolution during the first pulse by the Hamiltonian~\eqref{eq:H1}. We then sweep over a range of secondary pulse frequencies and phases and calculate the evolution using the Hamiltonian~\eqref{eq:H2}, after which one can calculate the ground state population.

The results for a system of transmons are shown in Fig.~\ref{fig:transmission_pulse_probe}(a), from which we can identify several transitions. First of all, we observe decreased ground state population at the dark state frequency. This occurs because the first Rabi pulse is imperfect and it leaves some of the population to the ground state~\cite{Zanner2021}, which the second pulse can excite with a suitable frequency and symmetry. The dark state transition vanishes from the spectrum with the antisymmetric drive at~$\phi=\pi$, since then the secondary pulse does not couple to the dark state, but instead it excites the shortly lived bright state~$\ket{B_4}$, so that the system ends up in a state it was in before the secondary pulse. In a slightly lower frequency we observe a transition with a large linewidth. This is actually caused by the transitions from the dark state $\ket{D_3}$ to the states~$\ket{W_{13}}$ and~$\ket{B_{14}}$, which are almost resonant. In the low frequency we observe the transition from the dark state to the state~$\ket{W_6}$. Noteworthy is that all the visible transitions are symmetric. We do not see the antisymmetric transitions to states~$\ket{W_5}$ and~$\ket{F_{12}}$, because they decay back to the dark state~$\ket{D_3}$ and thus do not alter the ground state population, whereas states~$\ket{W_6}$,~$\ket{W_{13}}$ and~$\ket{B_{14}}$ decay to the bright state~$\ket{B_{14}}$, which further decays to the ground state, see Fig.~\ref{fig:two_pairs_levels}. Experimentally this measurement was performed in Ref.~\cite{Zanner2021}.

In Fig.~\ref{fig:transmission_pulse_probe}(b) we show the same results for a system of qubits. There are two main differences compared to the transmon system. First, the state~$\ket{W_6}$ does not exist in a qubit system, and thus there are no states visible in low frequency. Second difference is that the states~$\ket{W_{13}}$ and~$\ket{B_{14}}$ occur at different frequencies in the qubit system than in the transmon one. In Fig.~\ref{fig:transmission_pulse_probe}(c) we for completeness show also the results for a harmonic system, for which only one spectral line is visible. This occurs because in harmonic systems the two-excitation bright state~$\ket{B_4}\otimes\ket{B_4}$ cannot be excited from the state~$\ket{D_3}$. One can, however, excite the symmetric state~$\ket{D_3}\otimes\ket{D_3}$, but since this is a dark state, it does not decay. Moreover, also the antisymmetric state~$\ket{D_3}\otimes\ket{B_4}$ can be excited, but since it decays back to the dark state~$\ket{D_3}$, it does not affect the ground state population.

\section{Conclusions}\label{sec:conclusions}
In this work we studied analytically and numerically an array of transmons interacting coherently with the electromagnetic field inside a rectangular waveguide. This interaction results in a long range coherent exchange interaction, as well as correlated decay, depending on the relative positions of the transmons inside the waveguide. Transmons are typically considered qubits, and properties of such two-level systems have already been widely explored in a waveguide setup~\cite{lalumiere2013, Mirhosseini2019, Albrecht2019, Masson2020}. Here we instead modeled transmons as anharmonic oscillators, which is a more accurate description of the device. The anharmonicity acts as a many-body interaction between bosonic excitations of transmons.

We found that in an array of harmonic oscillators, whose excitations are non-interacting bosons, the decay rates of the brightest states scale linearly with the number of excitations~$N$ and the system size~$L$ as~$\gamma N L$, as opposed to two-level system where the maximal decay rate is achieved with half filling. The anharmonicity of the transmon decreases the decay rates from the non-interacting system, but the behavior in large filling is closer to that of harmonic oscillators than qubits. However, unlike the system of harmonic oscillators, a transmon system can display a superradiant burst of emission, similarly as a qubit system.

We then focused on a smaller system of two pairs of transmons. The transmons forming a pair are coupled capacitively, but the pairs interact with each other only through the waveguide. Such systems are readily realizable also experimentally, and their effective separation inside the waveguide can be adjusted by flux tuning their energies. The level structure and symmetry properties of the system eigenstates were studied in detail. We also provided numerical analysis on different measurement schemes for probing the properties of the system. The two-pair system can be used for realizing a computational qubit~\cite{Zanner2021}, and in order to efficiently control the effective qubit, it is important to understand also the characteristics of the higher levels of the system, which are affected by the bosonic nature of transmons. Extension of the system to contain several 10's of transmons provides a platform for studying interacting many-body quantum systems in a collective environment~\cite{Gonzalez-Tudela2015,Albrecht2019}. Especially, disorder in transmon energies leads to many-body localization~\cite{Roushan17, Orell2019}, whose stability and impact on collective effects could be explored further~\cite{Celardo2013, Celardo2013_3D, Fayard2021}.

In this work the three-dimensional rectangular waveguide effectively behaves as an effective one-dimensional object. However, the two- or three-dimensionality can be restored by positioning the transmons differently inside the waveguide. Further, in rectangular waveguide the propagation of radiation is restricted to frequencies above certain cutoff frequency. Here we mainly considered the case where all the transmons have been tuned far above the cutoff, so that its effect can be ignored. However, the group velocity of radiation inside the waveguide depends on the cutoff frequency, and as the frequency approaches the cutoff, the corresponding group velocity decreases. Thus, close to the cutoff, the dynamics of the environment can no longer be assumed to occur at much briefer time scales as those of the system, which leads to non-Markovian behavior. These systems can therefore provide an intriguing platform for studying also non-Markovian many-body physics.

\section*{Acknowledgements}
This research was financially supported by 
 the Emil Aaltonen Foundation, the Academy of Finland under grant nos.~316619~and~320086, the European Research Council (ERC) under the European Unions Horizon 2020 research and innovation program (714235), the Austrian Science Fund FWF within the DK-ALM (W1259-N27) the Austrian Science Fund FWF within the SFB-BeyondC (F7106-N38), and Canada First Research Excellence Fund. We also wish to acknowledge CSC--IT~Center~for~Science,~Finland for computational resources. Numerical calculations were done using C++ with linear algebra implemented from the Eigen library~\cite{eigenweb}.

\appendix

\section{Effective master equation for a transmon array in a rectangular waveguide}\label{sec:master}
In this section we provide detailed derivation of the master equation for a system of multilevel atoms (transmon array) inside a rectangular waveguide. We follow closely the derivation provided in Refs.~\cite{lalumiere2013, Lehmberg1970} for a 1D waveguide with the exception that the presence of the cutoff frequency for propagating waves is explicitly taken into account. The total Hamiltonian comprises the emitter system which is here the transmon array, the electromagnetic field of the waveguide and their interaction:
\begin{equation}
\hop_{\rm T} = \hop_{\rm sys} + \hop_{\rm F} + \hop_{\rm I}.  
\end{equation}
Assuming that the transmons are not coupled to each other we write their Hamiltonian as
\begin{equation}
    \hop_{\rm sys} = \sum_{mj}E_{mj}\spop^{mj}\smop^{mj},
\end{equation}
where~$\smop^{mj}$ annihilates~the ($m+1$)th state of the site~$j$,~$\smop^{mj} = \ket{m_j}\bra{(m+1)_j}$, and~$E_{mj}$ is the corresponding energy.

\subsection{Electromagnetic environment of the waveguide}\label{sec:EM_WG}
We assume that the waveguide is a rectangular metallic pipe whose width in the~$x$ direction is~$a$ and in the~$y$ direction~$b$. In these restricted dimensions only standing electromagnetic modes are supported. In the~$z$ direction, we assume that the waveguide is infinite. Along this dimension, two possible types of electromagnetic waves can propagate:
Transverse electric modes~(TE) are such that the electric field has no~$z$ component,~$\vec E = \begin{pmatrix}E_x&E_y&0\end{pmatrix}$. Transverse magnetic modes~(TM) on the other hand do not have parallel magnetic component,~$\vec B = \begin{pmatrix}B_x&B_y&0\end{pmatrix}$.

The electromagnetic field can be described in terms of the vector potential~$\vec A$ and the scalar potential~$V$ as
\begin{equation}
    \vec E = -\boldsymbol\nabla V -\frac{\partial\vec A}{\partial t},\quad    
    \vec B = \boldsymbol\nabla\times\vec A,
\end{equation}
and the behavior of the electromagnetic field is determined by Maxwell's equations, which can be written as wave equations for the electromagnetic potentials, 
\begin{align}
    \left(\nabla^2 -\frac{1}{c^2}\frac{\partial^2}{\partial t^2}\right)\vec A=\vec0,\quad
    \left(\nabla^2 -\frac{1}{c^2}\frac{\partial^2}{\partial t^2}\right)V=0.
\end{align}
We recover the solutions
\begin{align}
    A_x(x,y,z,t) &= A_{x_0}\cos\left(k_xx\right)
    \sin\left(k_yy\right)e^{i(k_z z_j -\omega t)},\\
    A_y(x,y,z,t) &= A_{y_0}\sin\left(k_xx\right)
    \cos\left(k_yy\right)e^{i(k_z z_j -\omega t)},\\
    A_z(x,y,z,t) &= A_{z_0}\sin\left(k_xx\right)
    \sin\left(k_yy\right)e^{i(k_z z_j -\omega t)},\\
    V(x,y,z,t) &= \frac{c^2k_z}{\omega}A_z(x,y,z,t),
\end{align}
where we have defined the frequency as~$\omega=ck$ with the wavenumber~$k=\sqrt{k_x^2+k_y^2+k_z^2}$ and the speed of light $c$. The wavenumber is discretized in the~$x$ and~$y$ directions (standing modes),
\begin{equation}
    k_x = \frac{\alpha\pi}{a},\quad k_y = \frac{\beta\pi}{b},
\end{equation}
where $\alpha,\beta\in\mathbb{N}$. From this we recover a dispersion relation for the propagating waves,
\begin{align}\label{eq:dispersion}
  \omega_{\alpha\beta}(k_z) &= \sqrt{c^2k_z^2 + \left(\frac{c\alpha\pi}{a}\right)^2 + \left(\frac{c\beta\pi}{b}\right)^2}\notag \\
              &=\sqrt{c^2k_z^2 + \Omega_{\perp,\alpha\beta}^2},
\end{align}
where~$\Omega_{\perp, \alpha\beta}$ is the so-called cutoff frequency. Radiation with frequency below this cannot propagate through the waveguide. From the dispersion relation we obtain the phase velocity
\begin{equation}
    v_{\alpha\beta, \rm p}(k_z) = \frac{\omega_{\alpha\beta}(k_z)}{k_z} = \frac{c}
    {\sqrt{1 - \frac{\Omega_{\perp,\alpha\beta}^2}{\omega_{\alpha\beta}^2(k_z)}}},
\end{equation}
and the group velocity
\begin{equation}
    v_{\alpha\beta, \rm g}(k_z) = \frac{d\omega_{\alpha\beta}(k_z)}{dk_z} 
    = c\sqrt{1-\frac{\Omega_{\perp,\alpha\beta}^2}{\omega_{\alpha\beta}^2(k_z)}}.
\end{equation}
From the group velocity we notice that as the frequency~$\omega_{\alpha\beta}$ approaches the cutoff frequency, the group velocity decreases. Non-Markovian effects start to emerge once the system lenght scale~$d$ becomes~$d\gtrsim v_{\rm g}/\gamma$~\cite{Laakso2014}. For the parameters used in this work this happens only very close to the cutoff frequency.

Following the standard quantization, we obtain the vector potential 
\begin{align}
        \hat{\vec A}(\vec r, t) = &\sum_{\alpha\beta}\int_{-\infty}^\infty 
        dk_z\sqrt{\frac{\hbar\mu_0 c^2}{2\omega_{\alpha\beta}(k_z)}}
        \label{eq:Aop}\\
        \times&\left[\aop_{\alpha\beta k_z}e^{-i\omega_{\alpha\beta}(k_z)}\vec R(\vec r) + \aop_{\alpha\beta k_z}^\dag e^{i\omega_{\alpha\beta}(k_z)}\vec R^*(\vec r)\right],\notag
\end{align}
where~$\aop_{\alpha\beta k_z}^\dag$ creates a quantum to the waveguide field with a wavenumber~$k=\sqrt{k_z^2 + \left(\alpha\pi/a\right)^2+ \left(\beta\pi/b\right)^2}$ and the spatial dependence is given through the vector
\begin{equation}
    \vec R(\vec r) = 
    \begin{pmatrix}
        A_{x_0}\cos\left(\frac{\pi\alpha}{a}x\right)
        \sin\left(\frac{\pi\beta}{b}y\right)e^{ik_z z}\\
        A_{y_0}\sin\left(\frac{\pi\alpha}{a}x\right)
        \cos\left(\frac{\pi\beta}{b}y\right)e^{ik_z z}\\
        A_{z_0}\sin\left(\frac{\pi\alpha}{a}x\right)
        \sin\left(\frac{\pi\beta}{b}y\right)e^{ik_z z}
    \end{pmatrix}. \label{eq:Rvec}
\end{equation}
For TE-modes we can set $A_z=V=0$, and for TM-modes $A_x=A_y=0$.

We assume that only the TE10 mode interacts with the system and thus we set~$\alpha=1$,~$\beta=0$ and define~$\Omega_{\perp,10} \equiv\Omega_\perp$ in Eqs.~\eqref{eq:dispersion},~\eqref{eq:Aop} and~\eqref{eq:Rvec}, resulting in the dispersion relation $\omega(k_z)=\sqrt{c^2k_z^2 + \Omega_{\perp}^2}$.
Now, we recover the electric field as $\hat{\vec{E}}(\vec r, t) = -\partial\hat{\vec{A}}/\partial t$ as
\begin{align}
        \hat{\vec{E}}(\vec r, t) =&i\int_{-\infty}^\infty 
        dk_z\sqrt{\frac{\hbar\omega(k_z)\mu_0c^2}{2}}\sin\left(\frac{\pi x}{a}\right) \\
        \times&\left[\aop_{k_z}e^{-i(\omega(k_z)t-k_zz)}-\aop_{k_z}^\dag e^{+i(\omega(k_z)t-k_zz)}\right] A_{y_0}\bm{y}. \notag 
\end{align}
The Hamiltonian is 
\begin{equation}
    \hop_{\rm F} = \hbar\int_{-\infty}^\infty dk_z \omega(k_z) \aop_{k_z}^\dag\aop^{}_{k_z}.
  \end{equation}
for the TE10 radiation field inside the rectangular waveguide.

\subsection{Coherent interaction with the electromagnetic environment of the waveguide}\label{sec:master_equation}
We assume bilinear coupling between the atoms and the electric field, giving the coupling Hamiltonian
\begin{equation}
    \hop_{\rm I} = \hbar\sum_{mj}g_j\sqrt{m+1}\left(\xiop_j+\xiop_j^\dag\right)
    \sxop^{mj},
\end{equation}
where the position operator is~$\sxop^{mj} = \spop^{mj} + \smop^{mj}$, the coupling strength for the~$j$th atom is denoted with~$g_j$, and the operator related to the electric field is
\begin{equation}\label{eq:field_op}
    \xiop_j = -ic\int_{-\infty}^\infty dk_z\sqrt{\omega(k_z)}
    \sin\left(\frac{\pi x_j}{a}\right)e^{ik_z z_j}\aop_{k_z},
\end{equation}
where~$z_j$ and~$x_j$ are the coordinates of the $j$th atom.

\subsection{Dynamics of the electromagnetic fields}\label{sec:dynamics_of_field}
By utilizing the full Hamiltonian $\hat H_{\rm T}=\hop_{\rm sys} + \hop_{\rm F} + \hop_{\rm I}$, the dynamics of the field operator~$\aop_{k_z}(t)$ are determined by the Heisenberg equation of motion 
\begin{align}
        \frac{\aop_{k_z}}{dt}&=\frac{i}{\hbar}[\hop_{\rm T}, \aop_{k_z}]= -i\omega(k_z)\aop_{k_z}\\
        &+ \sum_{mj}cg_j\sqrt{m+1}\sqrt{\omega(k_z)} \sin\left(\frac{\pi x_j}{a}\right)e^{-ik_z z_j}\sxop^{mj}, \notag 
\end{align}
which has the solution up to time~$t$ 
\begin{align}
        \aop_{k_z}(t) =& \aop_{k_z}(0)e^{-i\omega(k_z)t}\notag \\
        &+\sum_{mj}cg_j\sqrt{m+1}\sqrt{\omega(k_z)}
        \sin\left(\frac{\pi x_j}{a}\right)e^{-ik_z z_j} \notag \\
        &\quad \times \int_0^t d\tau e^{-i\omega(k_z)(t-\tau)}\sxop^{mj}(\tau), \label{eq:aop_sol}
\end{align}
where the latter part describes the interaction with the transmons. With this we can write Eq.~\eqref{eq:field_op} as
\begin{align}
        &\xiop_j(t) = \xiop_j^{\rm in}(t) -i\sum_{nk}cg_k\sqrt{n+1} \sin\left(\frac{\pi x_j}{a}\right) \sin\left(\frac{\pi x_k}{a}\right) \notag  \\
        &\times\int_{-\infty}^\infty dk_z\omega(k_z)e^{ik_z(z_j-z_k)} \int_0^t d\tau e^{i\omega(k_z)(\tau-t)}\sxop^{nk}(\tau),  \label{eq:field_op2}
\end{align}
where we have defined
\begin{align}
    \xiop_j^{\rm in}(t) = 
    \frac{c}{i}&\int_{-\infty}^\infty dk_z \label{eq:infield}\\
    &\times\sqrt{\omega(k_z)}
    \sin\left(\frac{\pi x_j}{a}\right)
    e^{i(k_z z_j-\omega(k_z)t)}\aop_{k_z}(0)\nonumber
\end{align}

Our next objective is to calculate the integrals in Eq.~\eqref{eq:field_op2}. We do the Markov approximation by assuming weak coupling between the atoms and the environment, so that we can approximate $\smop^{nk}(\tau)\approx e^{-i\omega_{nk}(\tau-t)}\smop^{nk}$, where~$\omega_{nk} = (E_{n+1,k} - E_{nk})/\hbar)$ is the transition frequency between the~$(n+1)$st and~$n$th eigenstates of the~$j$th transmon. We also assume that the dynamics in the environment occur at much faster rate than those in the system, so we can extend the integration limit to infinity in the time integral. This gives
\begin{widetext}
\begin{align}
    I_{nkj} &= \int_{-\infty}^\infty
    dk_z\omega(k_z)e^{ik_zz_{jk}}
    \int_0^\infty d\tau e^{-i\omega(k_z)(t-\tau)}\sxop^{nk}(\tau) \notag \\ \notag  & \approx \int_{-\infty}^\infty dk_ze^{ik_zz_{jk}}\frac{\omega(k_z)}{\omega_{nk}}\Bigg\{ \spop^{nk}\left[\pi\delta\left(\frac{\omega(k_z)+\omega_{nk}}{\omega_{nk}}\right)-i\pv\frac{\omega_{nk}}{\omega(k_z)+\omega_{nk}}\right]\\ &\qquad \qquad \qquad \qquad \qquad \quad  +\smop^{nk}\left[\pi\delta\left(\frac{\omega(k_z)-\omega_{nk}}{\omega_{nk}}\right)-i\pv\frac{\omega_{nk}}{\omega(k_z)-\omega_{nk}}\right]\Bigg\},
\end{align}
where~$\pv$ is the Cauchy principal value. Next, we convert the integration over the positive wavenumbers only and  change the integration over wavenumber to integration over frequency using the dispersion relation~$k_z = \sqrt{\omega^2(k_z)-\Omega_\perp^2}/c$. We obtain
\begin{align}
      I_{nkj} =&2\smop^{mj}\frac{\omega_{nk}^2}{c\sqrt{\omega_{nk}^2-\Omega_\perp^2}} \cos\left(t_{jk}\sqrt{\omega_{nk}^2-\Omega_\perp^2}\right)\Theta\left(\omega_{nk}-\Omega_\perp\right) \notag \\
      &-2i\pv\int_{\Omega_\perp}^\infty d\omega \left[\spop^{mj} \frac{\omega\cos\left(t_{jk}\sqrt{\omega^2-\Omega_\perp^2}\right)} {c\sqrt{\omega^2-\Omega_\perp^2}\left(\omega+\omega_{nk}\right)}+\smop^{mj} \frac{\omega\cos\left(t_{jk}\sqrt{\omega^2-\Omega_\perp^2}\right)}{c\sqrt{\omega^2-\Omega_\perp^2}\left(\omega-\omega_{nk}\right)}\right],
\end{align}
where we have defined the propagation time~$t_{jk}$ in empty space between sites~$j$ and~$k$ as~$t_{jk} = |z_j-z_k|/c$, and~$\Theta$ is the Heaviside step function. With this, Eq.~\eqref{eq:field_op2} becomes
\begin{equation}
        \xiop_j(t) = \xiop_j^{\rm in}(t)- \frac{1}{g_j}\sum_{nk}\left[W_{kj}^{n+}\spop^{nk} + \left(W_{kj}^{n-}+\frac{i\gamma_{kj}^n}{2}\right)\smop^{nk}\right], \label{eq:xit}
\end{equation}
where we have defined
\begin{align}
    \gamma_{kj}^n =& 4\pi g_jg_k\sqrt{n+1}
    \sin\left(\frac{\pi x_j}{a}\right)
    \sin\left(\frac{\pi x_k}{a}\right)    \Theta\left(\omega_{nk}-\Omega_\perp\right)\frac{\omega_{nk}^2}{\sqrt{\omega_{nk}^2-\Omega_\perp^2}}
    \cos\left(t_{jk}\sqrt{\omega_{nk}^2-\Omega_\perp^2}\right)\label{eq:gamma_kjn},\\
    W_{kj}^{n\pm} = & 2g_jg_k\sqrt{n+1} \sin\left(\frac{\pi x_j}{a}\right)\sin\left(\frac{\pi x_k}{a}\right)\pv\int_{\Omega_\perp}^\infty
    d\omega\frac{\omega^2\cos\left(t_{jk}\sqrt{\omega^2-\Omega_\perp^2}\right)}
    {\sqrt{\omega^2-\Omega_\perp^2}
    \left(\omega\pm\omega_{nk}\right)}. \label{eq:W_kjnpm}
\end{align}
\end{widetext}

\subsection{Master equation for the transmon array}\label{sec:master_equation_app}
We can then obtain the master equation for the reduced density operator of the transmon array system by first considering the time evolution of an arbitrary operator~$\oop$ acting on the transmon array system only. The Heisenberg equation of motion gives
\begin{align}
  &\frac{d\oop}{dt} = \frac{i}{\hbar}\left[\hop_{\rm sys} + \hbar\sum_{mj}g_j\sqrt{m+1}\left(\xiop_j^{\rm in}+ \xiop_j^{\rm in\dag}\right)\sxop^{mj},\oop\right] \notag \\
                   &-i\sum_{mj,nk}\sqrt{m+1} \notag \\
        &\times\Bigg[W_{kj}^{n+}\left(\smop^{mj}\oop\spop^{nk}-\oop\smop^{mj}\spop^{nk}-\smop^{nk}\oop\spop^{mk}+\smop^{nk}\spop^{mj}\oop\right)\notag \\
        &+W_{kj}^{n-}\left(\spop^{mj}\oop\smop^{nk}-\oop\spop^{mj}\smop^{nk}-\spop^{nk}\oop\smop^{mk}+\spop^{nk}\smop^{mj}\oop\right) \notag \\
        &+\frac{\gamma_{kj}^n}{2}\left(\spop^{mj}\oop\smop^{nk}-\oop\spop^{mj}\smop^{nk}
        +\spop^{nk}\oop\smop^{mk}-\spop^{nk}\smop^{mj}\oop\right)\Bigg],\notag 
\end{align}
where we have performed the rotating wave approximation in terms of the type~$\sxop^{mj}\oop\spop^{nk}\approx\smop^{mj}\oop\spop^{nk}$. By using the fact that $\Tr_{\rm tot}\left(\frac{d\oop}{dt}\dens_{\rm tot}\right)=
\Tr\left(\oop\frac{d\dens}{dt}\right)$, where~$\Tr_{\rm tot}$ and~$\Tr$ are traces over total systems and transmons, respectively. Re-arranging the terms gives an equation of motion for the density matrix of the transmons in terms of 
the familiar Lindbladian dissipators, 
\begin{align}
        \frac{d\dens}{dt} = &-\frac{i}{\hbar}\left[\hop_{\rm sys}
          +\hbar\sum_{mj}L_{mj}\ket{m_j}\bra{m_j}, \dens\right]            \label{eq:master_raw}        \\
          &-i\left[\sum_{mj,nk}J_{mj,nk}\spop^{nk}\smop^{mj}
        +\sum_{mj}d_{mj}(t)\sxop^{mj}, \dens\right] \notag \\
        &+\sum_{mj,nk}\gamma_{mj,nk}\left(\smop^{mj}\dens\spop^{nk}-\frac{1}{2}\left\{\spop^{nk}\smop^{mj},\dens\right\}\right) \notag \\
+\sum_{mj,nk}&W_{mj,nk}\left(\spop^{mj}\dens\smop^{nk}+\smop^{nk}\dens\spop^{mj}-\left\{\smop^{nk}\spop^{mj},\dens\right\}\right), \notag 
\end{align}
where we have defined the radiation field induced driving as
\begin{equation}
    d_{mj}(t) = g_j\sqrt{m+1}\left[\braket{\xiop_j^{\rm in}(t)}
        +\braket{\xiop_j^{\rm in}(t)}^*\right], \label{eq:driving_me_raw}
\end{equation}
and the waveguide mediated exchange interaction  $J_{mj, nk}$ and the correlated decay coefficients $\gamma_{mj, nk}$ as
\begin{align}
    J_{mj,nk} =& \frac{i}{2}\Big(\sqrt{m+1}\frac{\gamma_{kj}^n}{2}-\sqrt{n+1}\frac{\gamma_{jk}^m}{2} \notag \\
                                &\qquad +i\sqrt{m+1}\widetilde W_{kj}^n +i\sqrt{n+1}\widetilde W_{jk}^m\Big),\\
  \gamma_{mj,nk} = &\sqrt{m+1}\frac{\gamma_{kj}^n}{2} +\sqrt{n+1}\frac{\gamma_{jk}^m}{2} \notag \\
    & \qquad +i\sqrt{m+1}\widetilde W_{kj}^n-i\sqrt{n+1}\widetilde W_{jk}^m,                    
\end{align}
with the shorthand notations
\begin{align}
  W_{mj,nk} &= i\left(\sqrt{m+1}W_{kj}^{n+}-\sqrt{n+1}W_{jk}^{m+}\right), \label{eq:mystery_W} \\
 \widetilde W_{jk}^n & = W_{jk}^{n+}+W_{jk}^{n-}. &\label{eq:widetilde_W}
\end{align}
The Lamb shift is
\begin{equation}\label{eq:lamb_shift}
    L_{mj} = \sqrt{m}W_{jj}^{(m-1)+}-\sqrt{m+1}W_{jj}^{m+},
\end{equation}
What then remains is to calculate expressions for the various coefficients in the master equation~\eqref{eq:master_raw}. 

\subsection{Above and below the cutoff frequency}\label{sec:above_and_below}
Next we compute the remaining master equation coefficients by paying attention to the cutoff frequency in the electromagnetic spectrum of the propagating modes in the waveguide.~Coefficient~$\gamma_{kj}^{n}$ was already calculated in~Eq.~\eqref{eq:gamma_kjn}.~For the principal value integral in Eq.~\eqref{eq:W_kjnpm} we obtain, after making a change of variables~$x=\sqrt{\omega^2-\Omega_\perp^2}$ and reordering,
\begin{align}
        I_{\pm} = &\omega_{nk}^2\pv\int_0^\infty dx\frac{\cos(t_{jk} x)}
        {x^2+\Omega_\perp^2-\omega_{nk}^2} \notag \\
        &\mp\omega_{nk}\int_0^\infty dx\frac{\sqrt{x^2+\Omega_\perp^2}\cos(t_{jk} x)}{x^2+\Omega_\perp^2-\omega_{nk}^2},\label{eq:difficult_integral}
\end{align}
where we have used the fact that $\int_0^\infty dx\cos(t_{jk} x) = 0$~\cite{lalumiere2013}. We have managed to divide the integral into two parts, one that can readily be calculated analytically:
\begin{equation}
    \int_0^\infty \frac{dx\cos(t_{jk}x)}{x^2+\Omega_\perp^2-\omega_{nk}^2}
    = \frac{\pi}{2}
    \begin{cases}
        \frac{e^{-t_{jk}\sqrt{\Omega_\perp^2-\omega_{nk}^2}}}{\sqrt{\Omega_\perp^2-\omega_{nk}^2}}
        ,&\omega_{nk}<\Omega_\perp\\
        -\frac{\sin\left(t_{jk}\sqrt{\omega_{nk}^2-\Omega_\perp^2}\right)}{\sqrt{\omega_{nk}^2-\Omega_\perp^2}}
         ,&\omega_{nk}>\Omega_\perp\\
        \infty, & \omega_{nk}=\Omega_\perp
    \end{cases}
\end{equation}
so that we have above the cutoff  $\omega_{nk}>\Omega_\perp$
\begin{align}
        \widetilde W_{kj}^n = -2&\pi g_jg_k\sqrt{n+1}\sin\left(\frac{\pi x_j}{a}\right)\sin\left(\frac{\pi x_k}{a}\right) \notag \\
        &\times \frac{\omega_{nk}^2}{\sqrt{\omega_{nk}^2-\Omega_\perp^2}}\sin\left(t_{jk}\sqrt{\omega_{nk}^2-\Omega_\perp^2}\right), 
\end{align}
and below the cutoff  $\omega_{nk}<\Omega_\perp$
\begin{align}
        \widetilde W_{kj}^n = 2&\pi g_jg_k\sqrt{n+1}\sin\left(\frac{\pi x_j}{a}\right)\sin\left(\frac{\pi x_k}{a}\right) \notag \\
        &\times\frac{\omega_{nk}^2}{\sqrt{\Omega_\perp^2-\omega_{nk}^2}}e^{-t_{jk}\sqrt{\Omega_\perp^2-\omega_{nk}^2}}.
\end{align}
The second integral in Eq.~\eqref{eq:difficult_integral} is much more difficult. However, they cancel in Eq.~\eqref{eq:widetilde_W}, and thus do not affect the correlated decay and exchange interaction terms, which above the cutoff are written as
\begin{align}
    \gamma_{mj,nk} =& 2\pi g_jg_k\sqrt{(m+1)(n+1)} \notag \\
    \times&\sin\left(\frac{\pi x_j}{a}\right)
    \sin\left(\frac{\pi x_k}{a}\right)
    \left(\chi_{mjk}+\chi_{nkj}^*\right),\\
    J_{mj,nk} =&-i\pi g_jg_k\sqrt{(m+1)(n+1)} \notag\\
    \times&\sin\left(\frac{\pi x_j}{a}\right)
    \sin\left(\frac{\pi x_k}{a}\right)
    \left(\chi_{mjk}-\chi_{nkj}^*\right),
\end{align}
where we have defined an oscillatory coefficient  
\begin{equation}
    \chi_{mjk} = \frac{\omega_{mj}^2}
    {\sqrt{\omega_{mj}^2-\Omega_\perp^2}}
    e^{it_{jk}\sqrt{\omega_{mj}^2-\Omega_\perp^2}}.
\end{equation}
Below the cutoff, we find similarly
\begin{align}
    \gamma^\perp_{mj,nk} =& -2i\pi g_jg_k\sqrt{(m+1)(n+1)} \notag\\
    \times & \sin\left(\frac{\pi x_j}{a}\right)
    \sin\left(\frac{\pi x_k}{a}\right)
    \left(\zeta_{mjk}-\zeta_{nkj}\right),\\
    J^\perp_{mj,nk} =&-\pi g_jg_k\sqrt{(m+1)(n+1)} \notag \\
    \times & \sin\left(\frac{\pi x_j}{a}\right)
    \sin\left(\frac{\pi x_k}{a}\right)
    \left(\zeta_{mjk}+\zeta_{nkj}\right),
\end{align}
with a coefficient that is exponentially decaying with the site separation~$t_{jk}$,
\begin{equation}
    \zeta_{mjk} = \frac{\omega_{mj}^2}
    {\sqrt{\omega_{mj}^2-\Omega_\perp^2}}
    e^{-t_{jk}\sqrt{\Omega_\perp^2-\omega_{mj}^2}}.
\end{equation}
Note that below the cutoff frequency the matrix~$\gamma_{mj,nk}$ is a traceless Hermitian matrix. Thus, it is not semipositive, and the master equation is no longer of the Lindbladian form. However, since the system frequencies are close to each other,~$\gamma_{mj,nk}$ are small and can be neglected. Physical justification for this is that the dissipation in this setup occurs if the emitted photons propagate along the waveguide to infinity, which is not possible if the transmons emit with a frequency below the cutoff. However, the photons can still travel to nearby sites, which is seen as the coherent exchange interaction.

Next we calculate the driving terms of Eq.~\eqref{eq:driving_me_raw}. The operator~$\xiop_j^{\rm in}(t)$ of Eq.~\eqref{eq:infield} is separated into left and right moving parts,
\begin{align}
  \xiop_j^{\rm in}(t)
    =& -i\sin\left(\frac{\pi x_j}{a}\right)\int_{\Omega_\perp}^\infty
    d\omega\frac{\sqrt{\omega^3}}
    {\sqrt{\omega^2-\Omega_\perp^2}}e^{-i\omega t}\\
    &\times\left[e^{it_j\sqrt{\omega^2-\Omega_\perp^2}}
    \aop_{\rm R}(\omega) +e^{-it_j\sqrt{\omega^2-\Omega_\perp^2}}
    \aop_{\rm L}(\omega)\right]\nonumber.
\end{align}
Assuming that the system is driven with a frequency~$\omega_{\rm d}$ with a coherent state~$\ket{\{\alpha\}}$, such that
\begin{equation}\label{eq:coherent} 
    \aop_{\rm R/L}(\omega)\ket{\{\alpha\}} = 
    \sqrt{\frac{2\pi P_{\rm R/L}}{\hbar\omega_{\rm d}}}
    \delta(\omega-\omega_{\rm d})\ket{\{\alpha\}},
\end{equation}
so that the amplitude driving the system (transmon array) is 
\begin{align}
        d_{mj}(t) =& -\frac{2g_j\omega_d\sqrt{2\pi(m+1)}}
        {\sqrt{\hbar}\sqrt{\omega_d^2-\Omega_\perp^2}}
        \sin\left(\frac{\pi x_j}{a}\right)
        \Theta(\omega_{\rm d}-\Omega_\perp) \notag \\
        &\times\bigg[\sqrt{P_{\rm R}}
        \sin
        \left(\omega_{\rm d}t-t_j\sqrt{\omega_{\rm d}^2
        -\Omega_\perp^2}\right) \notag \\
       & \quad +\sqrt{P_{\rm L}}
        \sin
        \left(\omega_{\rm d}t+t_j\sqrt{\omega_{\rm d}^2
        -\Omega_\perp^2}\right)\bigg],
\end{align}
which we can write in terms of~$\gamma_{mj,mj}$ as
\begin{align}
        d_{mj}(t) =& -2\sqrt{\frac{\gamma_{mj,mj}}
        {2\hbar\omega_{mj}}}
        \sqrt{\frac{\omega_{\rm d}^2
        \sqrt{\omega_{mj}^2-\Omega_\perp^2}}
        {(\omega_{\rm d}^2-\Omega_\perp^2)\omega_{mj}}}
        \Theta(\omega_{\rm d}-\Omega_\perp)\notag \\
        &\times\bigg[\sqrt{P_{\rm R}}
        \sin\left(\omega_{\rm d}t-t_j
        \sqrt{\omega_d^2-\Omega_\perp^2}\right) \notag \\
        &\quad + \sqrt{P_{\rm L}}
        \sin\left(\omega_{\rm d}t+t_j\sqrt{\omega_{\rm d}^2
        -\Omega_\perp^2}\right)\bigg].
\end{align}
The system thus cannot be driven with a frequency below the cutoff, since such modes cannot propagate through the waveguide.

In Eqs.~\eqref{eq:mystery_W} and~\eqref{eq:lamb_shift} one is required to calculate the coefficient~$W_{kj}^{n+}$. The Lamb shift can be absorbed to the definition of the system frequencies~\cite{lalumiere2013}. Further, the matrix~$W_{mj,nk}$ is traceless and Hermitian, meaning it is not semipositive. Thus, the master equation is not of the Lindblad form. However, as shown in~Ref.~\cite{lalumiere2013}, and supported by numerical calculations, the actual values for~$W_{mj,nk}$ are in general small, and can be neglected. Thus, we obtain the master equation~\cite{lalumiere2013, Gu2017}
\begin{align}
        \frac{d\dens}{dt} = &-i\bigg[\frac{\hop_{\rm sys}}{\hbar} 
        + \sum_{mj,nk} J_{mj,nk}\spop^{nk}\smop^{mj},\dens \bigg] \notag \\
        & +\sum_{mj,nk}\gamma_{mj,nk}
        \left(\smop^{mj}\dens\spop^{nk}-\frac{1}{2}
          \left\{\spop^{nk}\smop^{mj},\dens\right\}\right)\notag \\
          &-i\bigg[\sum_{mj} d_{mj}(t)\sxop^{mj}, \dens\bigg] \label{eq:master_final}
\end{align}
In the main text in Secs.~\ref{sec:collective}-\ref{sec:conclusions}, we assume that the system frequencies~$\omega_{mj}$ are all well above the cutoff frequency~$\Omega_\perp$ so that we can effectively set~$\Omega_\perp=0$ for simplicity, and the coefficients reduce to those obtained in Ref.~\cite{lalumiere2013}.

\subsection{Input-output theory}\label{sec:master_equation_io}
We finish this section by deriving the input-output theory for the system of transmons inside the waveguide. This gives us tools to study the transmission and emission of radiation, as discussed in Sec.~\ref{sec:spectroscopy}. 
In Eq.~\eqref{eq:aop_sol} we presented a formal solution for the equation of motion of~$\aop_{k_z}(t)$ before the radiation has interacted with the transmons. Similar solution for time evolution up to time $t_f$ after the interaction reads
\begin{align}\label{eq:aop_sol2}
        \aop_{k_z}(t) = & \aop_{k_z}(t_f)e^{-i\omega(k_z)t} \notag \\
        &-\sum_{mj}cg_j\sqrt{m+1}\sqrt{\omega(k_z)}
        \sin\left(\frac{\pi x_j}{a}\right)e^{-ik_z z_j} \notag \\
        &\times\int_t^{t_f} d\tau e^{-i\omega(k_z)(t-\tau)}\sxop^{mj}(\tau).
\end{align}
Adding Eqs.~\eqref{eq:aop_sol} and~\eqref{eq:aop_sol2} together, separating left and right moving modes and integrating over $k_z$ gives
\begin{align}
        &\aop_{\rm R/L}^{\rm out}(t) - \aop_{\rm R/L}^{\rm in}(t) \notag \\
        &\quad =\sum_{mj}
        \sin\left(\frac{\pi x_j}{a}\right)
        \frac{\omega_{mj}g_j\sqrt{m+1}\sqrt{2\pi\omega_{mj}}}
        {\sqrt{\omega_{mj}^2-\Omega_\perp^2}}\notag \\
        &\qquad \quad \times e^{\mp it_j\sqrt{\omega_{mj}^2-\Omega_\perp^2}}
        \Theta(\omega_{mj}-\Omega_\perp)
        \smop^{mj}(t), \label{eq:aout1}
\end{align}
where we extended the integration limits in the time integral from~$-\infty$ to~$+\infty$, and defined
\begin{align}
    \aop_{\rm R/L}^{\rm in}(t) &= \frac{1}{\sqrt{2\pi}}
    \int_0^\infty dk_ze^{-i\omega(k_z)t}
    \aop_{\rm R/L}(\omega(k_z), 0),\\
    \aop_{\rm R/L}^{\rm out}(t) &= \frac{1}{\sqrt{2\pi}}
    \int_0^\infty dk_ze^{-i\omega(k_z)t}
    \aop_{\rm R/L}(\omega(k_z), t_f).
\end{align}
We can write Eq.~\eqref{eq:aout1} in terms of~$\gamma_{mj,mj}$ as
\begin{align}
        \aop_{\rm R/L}^{\rm out}(t) &-\aop_{\rm R/L}^{\rm in}(t)  \notag \\
        &=\sum_{mj}\sqrt{\frac{\gamma_{mj,mj}}{2}}
        \sqrt{\frac{\omega_{mj}}
        {\sqrt{\omega_{mj}^2-\Omega_\perp^2}}}\notag \\
        &\quad \times e^{\mp it_j\sqrt{\omega_{mj}^2-\Omega_\perp^2}}
        \Theta(\omega_{mj}-\Omega_\perp).
        \smop^{mj}(t)
\end{align}
The expectation value~$\braket{\aop_{\rm R/L}^{\rm in}}$ is obtained using Eq.~\eqref{eq:coherent}:
\begin{equation}
   \braket{\aop_{\rm R/L}^{\rm in}(t)} 
   = \frac{1}{c}\sqrt{\frac{P_{\rm R/L}}{\hbar}}
   \sqrt{\frac{\omega_{\rm d}}{\omega_{\rm d}^2-\Omega_\perp^2}}
   e^{-i\omega_{\rm d}t}\Theta(\omega_{\rm d}-\Omega_\perp).
\end{equation}
In the main text of Secs.~\ref{sec:collective}-\ref{sec:conclusions} we set~$\Omega_\perp=0$ because all the system frequencies are sufficiently far above the cutoff frequency. Assuming that the system is driven from the left only, we recover the input field
\begin{equation}\label{eq:ain}
     \braket{\aop_{\rm L}^{\rm in}(t)} 
     = \sqrt{\frac{P_{\rm L}}{\hbar\omega_{\rm d}}}
     e^{-i\omega_{\rm d}t},
\end{equation}
and the output field
\begin{equation}\label{eq:aout}
     \braket{\aop_{\rm L}^{\rm out}(t)} 
     = \braket{\aop_{\rm L}^{\rm in}(t)} 
     +\sum_{mj}e^{it_j\omega_{mj}}
     \sqrt{\frac{\gamma_{mj,mj}}{2}}\braket{\smop^{mj}(t)}.
\end{equation}
The transmission is defined as their ratio
\begin{equation}\label{eq:transmission}
     |t|^2 = \left|\frac{\braket{\aop_{\rm L}^{\rm out}(t)}}
     {\braket{\aop_{\rm L}^{\rm in}(t)}}\right|^2.
\end{equation}

\section{Non-Hermitian quantum mechanics}\label{sec:nonherm}
In standard quantum mechanics, observables are described by Hermitian operators with orthonormal eigenstates and real eigenvalues. Especially the Hermiticity of the Hamiltonian is required for the conservation of energy. 
However, realistic systems are in general non-conservative due to loss of particles, energy and information. These phenomena can be described with non-Hermitian Hamiltonians~\cite{Brody_2013, Ashida2020}, see Eqs.~\eqref{eq:gen_WQED_2}-\eqref{eq:effective_hamiltonian}. Consider a non-Hermitian Hamiltonian of the form
\begin{equation}\label{eq:bi01}
    \hop = \hop_{\rm R} - \frac{i}{2}\hop_{\rm I},
\end{equation}
with $\hop_{\rm R} = \hop_{\rm R}^\dag$ and $\hop_{\rm I} = \hop_{\rm I}^\dag$. Clearly $\hop \neq \hop^\dag$. The Hamiltonian $\hop$ has eigenvalues and eigenvectors
\begin{equation}
    \hop\ket{\alpha} = \lambda_\alpha\ket{\alpha},\quad
    \bra{\alpha}\hop^\dag = \bra{\alpha}\lambda_\alpha^*,
\end{equation}
where $\bra{\alpha} = \ket{\alpha}^\dag$, and the eigenvalues are of the form
\begin{equation}\label{eq:complex_energy}
    \lambda_\alpha = E_\alpha - i\hbar \frac{\Gamma_\alpha}{2},
\end{equation}
where we treat $E_\alpha$ as the energy and $\Gamma_\alpha$ as the decay rate of the state $\ket{\alpha}$. One can also calculate the eigenvalues of the Hermitian conjugate $\hop^\dag$:
\begin{equation}
    \hop^\dag\ket{\widetilde\alpha} = \widetilde \lambda_\alpha\ket{\widetilde \alpha},\quad
    \bra{\widetilde\alpha}\hop = \bra{\widetilde\alpha}\widetilde \lambda_\alpha^*.
\end{equation}
The eigenstates $\{\ket{\alpha}\}$ are called \textit{right} eigenvectors, and $\{\ket{\widetilde\alpha}\}$ are called \textit{left} eigenvectors. Now, in general, the eigenvectors $\{\ket{\alpha}\}$ do not form an orthogonal set, i.e.,~it can occur that $\braket{\beta|\alpha} \neq0$ for $\beta \neq\alpha$. However, together with the conjugate basis $\{\ket{\widetilde\alpha}\}$ they form a biorthogonal basis~\cite{Brody_2013},
\begin{equation}
    \braket{\widetilde\beta|\alpha} = 
    \delta_{\beta\alpha}\braket{\widetilde\alpha|\alpha},
\end{equation}
and $\braket{\widetilde\alpha|\alpha}\neq0$. Note that even though the states are biorthogonal, they are not necessarily orthonormal. Thus, the identity operator in this biorthogonal basis takes the form 
\begin{equation}
    \hat I = \sum_{\alpha}\frac{\ket{\alpha}\bra{\widetilde\alpha}}
    {\braket{\widetilde\alpha|\alpha}},
\end{equation}
where the denominator ensures that $\hat I^2=\hat I$.

\subsection{Expectation values and decay channels}
The biorthogonal basis changes the definitions of inner products and expectation values. Assume we have a general state $\ket{\psi}$, which we can write as a linear combination of either right or left eigenvectors,
\begin{align}
    \ket{\psi} & = \sum_\alpha \psi_\alpha\ket{\alpha}, & \psi_\alpha & = \frac{\braket{\widetilde\alpha|\psi}}{\braket{\widetilde\alpha|\alpha}},\\
     \ket{\widetilde\psi} & = \sum_\beta\widetilde\psi_\beta\ket{\widetilde\beta}, & \widetilde\psi_\beta &= \frac{\braket{\beta|\widetilde\psi}}{\braket{\beta|\widetilde\beta}}.
\end{align}
With these, the inner product between two arbitrary states $\ket{\psi}$ and $\ket{\phi}$ becomes
\begin{align}
    \braket{\widetilde\phi|\psi} &= \sum_{\alpha\beta}\widetilde\phi_\beta^*
    \psi_\alpha\braket{\widetilde\beta|\alpha} 
    = \sum_\alpha\frac{\braket{\widetilde\phi|\alpha}
    \braket{\widetilde\alpha|\psi}}{\braket{\widetilde\alpha|\alpha}}.
\end{align}
We define the expectation value of an arbitrary operator $\hat A$ in state $\ket{\phi}$ analogously as
\begin{equation}
    \braket{\hat A} = 
    \frac{\braket{\widetilde\phi|\hat A|\phi}}{\braket{\widetilde\phi|\phi}},
\end{equation}
and as a special case, the expectation value in an eigenstate of a non-Hermitian Hamiltonian is
\begin{equation}
    \braket{\hat A}_\beta = \frac{\braket{\widetilde\beta|\hat A|\beta}}{\braket{\widetilde\beta|\beta}}.
\end{equation}
The non-Hermitian Hamiltonian can have $m$-fold degenerate eigenstates, i.e., an identical complex eigenvalue for several states
\begin{equation}
    \lambda_\alpha = \frac{\braket{\widetilde\alpha_i|\hop|\alpha_i}}{\braket{\widetilde\alpha_i|\alpha_i}},\quad
    i = 1,2,\dots,m.
\end{equation}
In such cases the numerical diagonalization might not give the correct biorthogonal eigenstates, but one instead has to biorthogonalize them separately by using e.g.,~the Gram--Schmidt process. New right and left eigenvectors can be obtained with the modified algorithm as
\begin{alignat}{2}
    \ket{\phi_\alpha^k} &= \ket{\alpha_k} - \sum_{j=1}^{k-1}
    \frac{\braket{\widetilde{\phi}_\alpha^j|\alpha_k}}
    {\braket{\widetilde{\phi}_\alpha^j|\phi_\alpha^j}}
    \ket{\phi_\alpha^j},\\
    \ket{\widetilde{\phi}_\alpha^k} &= \ket{\widetilde{\alpha}_k}
    - \sum_{j=1}^{k-1}
    \frac{\braket{\phi_\alpha^j|\widetilde{\alpha}_k}}
    {\braket{\phi_\alpha^j|\widetilde{\phi}_\alpha^j}}
    \ket{\widetilde{\phi}_\alpha^j},
\end{alignat}
where we start with $\ket{\phi_\alpha^1} = \ket{\alpha_1}$ and
$\ket{\widetilde{\phi}_\alpha^1} = \ket{\widetilde{\alpha}_1}$.

Once we have obtained the eigenstates of the effective Hamiltonian, we can calculate the decay channels, that is the decay rates between the states induced by the jump operators of the master equation. The total decay rate of a state is given by the imaginary part of the respective eigenvalue,
\begin{equation}
    \Gamma_\alpha = -\frac{2}{\hbar}\im\left(\frac{\braket{\widetilde{\alpha}|\hop|\alpha}}
    {\braket{\widetilde{\alpha}|\alpha}}\right).
\end{equation}
Starting from Eqs.~\eqref{eq:bi01}-\eqref{eq:complex_energy}
\begin{align}
    &\braket{\alpha|\hop_{\rm R}|\beta} - \frac{i}{2}\braket{\alpha|\hop_{\rm I}|\beta}
    =\lambda_\beta\braket{\alpha|\beta},\\
    &\braket{\alpha|\hop_{\rm R}|\beta} + \frac{i}{2}\braket{\alpha|\hop_{\rm I}|\beta}
    = \lambda_\alpha^*\braket{\alpha|\beta},
\end{align}
we obtain
\begin{equation}
    \braket{\alpha|\beta} 
    = 2\frac{\braket{\alpha|\hop_{\rm R}|\beta}}
    {\lambda_\alpha^*+\lambda_\beta}
    = i\frac{\braket{\alpha|\hop_{\rm I}|\beta}}
    {\lambda_\alpha^*-\lambda_\beta}.
\end{equation}
Setting $\beta = \alpha$ we obtain
\begin{equation}
    E_\alpha = \frac{\braket{\alpha|\hop_{\rm R}|\alpha}}
    {\braket{\alpha|\alpha}},\quad
    \Gamma_\alpha = \frac{\braket{\alpha|\hop_{\rm I}/\hbar|\alpha}}
    {\braket{\alpha|\alpha}},
\end{equation}
where we have used Eq.~\eqref{eq:complex_energy}.
Further, the imaginary part of the Hamiltonian can be written as
\begin{equation}
    \hop_{\rm I} =\hbar  \sum_k \gamma_k\bop_k^\dag\bop^{}_k,
\end{equation}
where~$\bop_k$ are the jump operators, and~$\gamma_k$ gives the jump rates. Using this we obtain an expression for the total decay rate
\begin{align}
    \Gamma_\alpha &= 
    \sum_k\gamma_k\frac{\braket{\alpha|\bop_k^\dag\bop^{}_k|\alpha}}
    {\braket{\alpha|\alpha}} \notag \\
    &=\sum_k\frac{\gamma_k}{\braket{\alpha|\alpha}}
    \left\langle \alpha \left|\bop_k^\dag\sum_\beta\frac{\ket{\beta}\bra{\widetilde\beta}}
    {\braket{\widetilde\beta|\beta}}\bop_k \right|\alpha \right\rangle, \notag \\
    &=\sum_\beta\sum_k\gamma_k
    \frac{\braket{\alpha|\bop_k^\dag|\beta}
    \braket{\widetilde\beta|\bop_k|\alpha}}
    {\braket{\alpha|\alpha}\braket{\widetilde\beta|\beta}},
\end{align}
where we recover that the decay rate caused by the $k$th jump operator from the state $\ket{\alpha}$ to the state $\ket{\beta}$ is 
\begin{equation}
    \Gamma_{\alpha\to\beta}^k = 
    \gamma_k
    \frac{\braket{\alpha|\bop_k^\dag|\beta}
    \braket{\widetilde\beta|\bop_k|\alpha}}
    {\braket{\alpha|\alpha}\braket{\widetilde\beta|\beta}}.
\end{equation}
In the case of a Hermitian system, the result reduces to Fermi's golden rule,~$\Gamma_{\alpha\to\beta}^k=\gamma_k|\braket{\beta|\bop_k|\alpha}|^2$.

\section{Numerical time evolution}\label{sec:num_evol}\label{sec:numerics}
The unitary time evolution of an open quantum system is governed by a master equation, such as Eq.~\eqref{eq:gen_WQED}, which we can write in the form of
\begin{equation}
    \frac{d\dens}{dt} = \lind(t)\dens,
\end{equation}
where $\lind$ is the Liouvillian superoperator, and $\dens$ is the system density operator. In numerical calculations we first transform the operators and superoperators in the master equation into vectors and matrices, respectively. Suppose that the dimension of the Hilbert space is $d$. Then, the density operator is a $d\times d$ -dimensional matrix, which we tweak into a $1\times d^2$ column vector $\vec r$ by stacking the columns of $\hat \rho$ on top of each other. 
The products between the operators and the density operator then change to matrix vector products~\cite{AmShallem2015},
\begin{align}
    \hat A\dens\hat B^\dag \to \Big((\mat B^\dag)^T \otimes \mat A\Big)\vec r, \label{eq:op_to_vect}
\end{align}
where~$\mat A$ and $\mat B$ are the~$d\times d$ matrix forms of the operators~$\hat A$ and~$\mat B$. One sided operations such as $\hat H \dens $ are understood by replacing one operator in Eq.~\eqref{eq:op_to_vect} by a~$d\times d$ identity matrix $\mat I$. With these one can write a master equation as a matrix-vector equation
\begin{equation}\label{eq:me_mat_vec}
    \frac{d\vec r}{dt} = \mat L(t)\vec r,
\end{equation}
which can be solved with conventional numerical methods.

For a time independent system the steady state density operator $\rhoss$ is defined as a state that does not change in time, i.e.
\begin{equation}\label{eq:ss}
    \frac{d\rhoss}{dt} = 0\implies\lind \rhoss = 0\implies \mat L \vec r_{\rm ss}=0
\end{equation}
In general, if the system contains dark states, the steady state is not unique and we can merely define a manifold of steady states. However, since we always include also the bulk dissipation, also dark states decay and there exists only one steady state.

\subsection{Time-independent Liouvillian}
If the Liouvillian is time-independent, then the time evolution generated by the master equation~\eqref{eq:me_mat_vec} is solved by
\begin{equation}
    \vec r(t) = e^{\mat Lt}\vec r_0,  \label{eq:time_evol_rL}
\end{equation}
with $\vec r_0$ the initial state of the system. If the system is small enough, one is able to diagonalize the Liouvillian~$\mat L$, in which case the matrix exponential is trivial. 

Full diagonalization is in many cases impractical as the dimension of the Liouvillian matrix increases as $d^2\times d^2$. The Krylov subspace method~\cite{saadIterativeMethods, molerNineteenDubiousWays2003,manmanaTimeEvolution,beerwerthKrylovSubspaceMethods2015, luitzErgodicSideManybody} that can be formulated to employ efficiently sparse matrices 
is sufficiently accurate and numerically affordable method for our purposes here. Assume that we know the state of the system $\vec r(t)$ at time~$t$. After a brief time $\Delta t$ the state becomes
\begin{equation}
    \vec r(t+ \Delta t ) = e^{\mat L \Delta t}\vec r(t).
\end{equation}
If the time step $\Delta t$ is sufficiently short, one can accurately express the states $\vec r(t)$ and the  Liouvillian matrix $\mat L$ in an $m$-dimensional subspace $\mathcal{K}_m$ where $m\ll d^2$. This subspace is spanned by the vectors 
\begin{equation}
    \big\{\vec v_0,\mat L\vec v_0, \mat L^2\vec v_0,\dots, \mat L^{m-1}\vec v_0\big\},
\end{equation}
where we have defined $\vec v_0 \equiv \vec r(t)$. This basis is not orthogonal, but one can construct an orthonormal basis with the Arnoldi iteration using the Gram--Schmidt process, which results in an orthonormal unitary matrix $\mat K_m$ constructed from the orthonormalized vectors
\begin{equation}
    \mat K_m = 
    \begin{pmatrix}
        \vec v_0 & \vec v_1 & \vec v_2 & \dots & \vec v_{m-1}
    \end{pmatrix},
\end{equation}
and an upper Hessenberg matrix $\mat M_m$, such that~\cite{saadIterativeMethods, molerNineteenDubiousWays2003}
\begin{equation}
    \mat K_m^\dag\mat L\mat K_m = \mat M_m.
\end{equation}
At each step of the Arnoldi iteration one multiplies the previous vector by $\mat L$ and orthonormalizes it with respect to the previous ones using the Gram--Schmidt process. Once the matrices have been constructed, one can calculate the approximate time evolution as
\begin{equation}
    \vec r(t+\Delta t) \approx \mat K_m e^{\Delta t \mat M_m}\mat K_m^\dag \vec r(t),
\end{equation}
where the matrix exponential of the small matrix $\mat M_m$ is easily calculated e.g.~with the exact diagonalization or the Padé approximation. The Krylov method gives accurate results because the eigenvalues of the upper Hessenberg matrix $\mat M_m$ approximate the eigenvalues of the Liouvillian matrix that are the most important for the dynamics during the current time step.

\subsection{Time-dependent Liouvillian}
If the Liouvillian is time-dependent, solving the master equation is not as simple, as it would involve a time-ordered integral if expressed in the form of Eq.~\eqref{eq:time_evol_rL}. To recover the form, we apply the Magnus expansion~\cite{lubich2002}, in which case the solution takes the form
\begin{equation}
    \vec r(t+\Delta t) = e^{\mat U(t+\Delta t,t)}\vec r(t).
\end{equation}
Here the matrix $\mat U(t,0)$ is given by the Magnus series 
\begin{equation}
	\begin{aligned}
		\mat U(t,0) &= \int_0^t d t_1\mat M(t_1) + \frac{1}{2}
		\int_0^t d t_1\int_0^{t_1}d t_2\left[\mat M(t_1),\mat M(t_2)\right]\\
		&+\frac{1}{6}\int_0^t d t_1\int_0^{t_1}d t_2\int_0^{t_2}d t_3
		\Big\{\big[\mat M(t_1),\left[\mat M(t_2), \mat M(t_3)\right]\big]\\
		&+\big[\mat M(t_3),\left[\mat M(t_2), \mat M(t_1)\right]\big]
		\Big\}+\dots.
	\end{aligned}
\end{equation}
Truncating the series gives
\begin{equation}
    \mat U(t+{\Delta t},t) = {\Delta t}\mat B_0 - (\Delta t)^2\big[\mat B_0,\mat B_1\big] + \mathcal{O}\left[(\Delta t)^5\right],
\end{equation}
where the matrices $\mat B_k$ are 
\begin{equation}
    \mat B_k(t) = \frac{1}{{\Delta t}^{k+1}}\int_{-\frac{\Delta t}{2}}^{\frac{\Delta t}{2}}
    \tau^k\mat L\left(t+\tau + \frac{{\Delta t}}{2}\right)d\tau.
\end{equation}
In our studies, we found that the best numerical performance was achieved by simply using the lowest order expansion
\begin{equation}
    \vec r(t+{\Delta t}) = e^{{\Delta t}\mat B_0(t)}\vec r(t).
\end{equation}
The matrix exponential can then be calculated either exactly or with the Krylov subspace method described above. Notice that even though the time-independent system might be small enough to be solved using exact matrix exponentiation, time-dependent case of the same size is much heavier since the matrix exponential has to be calculated at each time step. Thus, in time-dependent case the Krylov method offers benefits.


%

\end{document}